\documentclass[american,aps,pra,reprint,tightenlines,superscriptaddress,twocolumn,twoside,longbibliography]{revtex4-2}
\pdfoutput=1

\usepackage[utf8]{inputenc}

\usepackage[unicode=true,pdfusetitle, bookmarks=true,bookmarksnumbered=false,bookmarksopen=false, breaklinks=false,pdfborder={0 0 0},backref=false,colorlinks=false] {hyperref}
\hypersetup{ colorlinks,linkcolor=myurlcolor,citecolor=myurlcolor,urlcolor=myurlcolor}
\usepackage{braket,colortbl,amsthm,amsmath,amssymb,dsfont}
\definecolor{myurlcolor}{rgb}{0,0,0.7}
\usepackage{graphics,graphicx,relsize}
\usepackage{color}
\usepackage{XCharter}
\usepackage[T1]{fontenc}

\usepackage{enumitem}  

\setlist[itemize]{left=0pt, labelindent=0pt}

\usepackage{graphicx}
\usepackage{subcaption}
\usepackage{url}
\usepackage[font=small,labelfont=bf,justification=raggedright]{caption}

\usepackage{siunitx}
\usepackage{mathptmx}
\usepackage[left=1.6cm,right=1.6cm,top=2.45cm,bottom=2.45cm]{geometry}
\usepackage{bigints}
\usepackage{accents}

\usepackage{multirow}
\usepackage{multirow}
\usepackage[table,xcdraw]{xcolor}

\usepackage{float} 

\theoremstyle{plain}

\def\bea{\begin{eqnarray}}
\def\eea{\end{eqnarray}}
\def\ba{\begin{array}}
\def\ea{\end{array}}

\def\beq{\begin{equation}}
\def\eeq{\end{equation}}

\usepackage[normalem]{ulem}
 
\begin{document}
\begingroup
\maketitle
\endgroup

\title{Flat-band thermodynamics reveals enhanced performance across Otto, Carnot, and Stirling cycles}

\author{Hadi Mohammed Soufy}
\email{hm.soufy@niser.ac.in}
\affiliation{School of Physical Sciences, National Institute of Science Education and Research, HBNI, Jatni-752050, India}
\affiliation{Homi Bhabha National Institute, Training School Complex, Anushakti Nagar, Mumbai, 400094, India}

\author{Colin Benjamin}
\email{colin.nano@gmail.com}
\affiliation{School of Physical Sciences, National Institute of Science Education and Research, HBNI, Jatni-752050, India}
\affiliation{Homi Bhabha National Institute, Training School Complex, Anushakti Nagar, Mumbai, 400094, India}

\begin{abstract}
Magic-angle twisted bilayer graphene (MATBG) exhibits remarkable electronic properties under external magnetic fields, notably the emergence of flat Landau levels. In this study, we present a comprehensive analysis of MATBG’s operational phase diagram under three distinct quantum thermodynamic cycles, i.e., Quantum Otto Cycle (QOC), Quantum Carnot Cycle (QCC), and Quantum Stirling Cycle (QSC). Employing the continuum eight-band model, we evaluate the thermodynamic performance of MATBG across multiple operational modes: heat engine, refrigerator, cold pump, and Joule pump, and benchmark it against other graphene systems such as monolayer graphene, AB-Bernal stacked bilayer graphene, and non-magic-angle twisted bilayer graphene. Our findings reveal that MATBG demonstrates superior heat engine performance in QSC, while achieving high efficiency albeit with reduced work output in  QOC. Even though the performance of MATBG as a cold pump or refrigerator is modest in QOC and QSC, it shows notable improvement as a refrigerator in QCC. Additionally, we identify a highly reversible Joule pump mode in both QSC and QOC under strict adiabaticity, underscoring the unique thermodynamic behavior of MATBG.
\end{abstract}
\maketitle
\clearpage  

\section{Introduction}

Quantum thermodynamics investigates how classical thermodynamic principles emerge and are modified in quantum systems, where inherently quantum features, such as discrete energy spectra, coherence, and entanglement, lead to nonclassical behavior and novel thermodynamic effects \cite{gemmer2009quantum}. A central focus of this field is the study of quantum thermodynamic cycles, such as the Quantum Carnot Cycle (QCC), Quantum Otto Cycle (QOC), and Quantum Stirling Cycle (QSC), which adapt classical protocols (adiabatic, isothermal, and isochoric processes) to quantum systems for performing work or mediating heat exchange \cite{HTQuan,quan2009quantum}. These cycles, like their classical counterparts, can be classified into operational regimes such as heat engines, refrigerators, cold pumps, and Joule pumps, based on the directionality of heat and work flows \cite{prakash2022impurity,scharf2020topological}. However, the presence of quantum effects can significantly modify both the performance bounds and the physical implementation of these regimes. Quantum heat engines have been realized across a diverse array of experimental platforms, including spin systems \cite{peterson2019experimental}, atomic motional degrees of freedom \cite{bouton2021quantum}, graphene based devices \cite{pena2020quasistatic}, topological Josephson junctions \cite{scharf2020topological}, and minimal two-level systems \cite{HTQuan}. Similarly, quantum refrigerators and cold pumps that exploit quantum effects to control heat transport at microscopic scales have been demonstrated using nitrogen-vacancy centers and superconducting qubits \cite{zaiser2021cyclic}, trapped ions \cite{maslennikov2019quantum}, and two-stroke two-qubit machines driven by measurement-based protocols \cite{buffoni2019quantum}. Quantum analogs of Joule pumps have also been proposed and realized, notably in one-dimensional Josephson-junction loops with spin-flipping impurities \cite{pal2022josephson} and in topological superconducting systems \cite{scharf2020topological}. With rapid experimental progress, many of these theoretical proposals are now becoming experimentally accessible, and novel quantum thermal machines continue to emerge across a variety of physical platforms \cite{myers2022quantum}.

Twisted bilayer graphene (TBG) has emerged as a compelling platform for investigating a variety of quantum phenomena, including strongly correlated electronic phases \cite{elecPhase1,elecPhase2}, superconductivity \cite{supercond}, and topologically nontrivial states \cite{topo1,topo2}. At specific twist angles, most notably at the so-called magic angle $\theta_{\text{twist}} \sim 1.05^\circ$, the resulting moiré superlattice gives rise to flat electronic bands, which dramatically enhance electron-electron interactions and foster emergent correlated behaviors \cite{moire}. Under a perpendicular magnetic field, TBG exhibits a highly tunable Landau level spectrum, whose features can be precisely modulated by the twist angle and external gating \cite{wang2012fractal,zhang2019landau,topo1}. This high degree of tunability positions TBG as a promising platform for exploring quantum thermodynamic processes, where external control parameters can directly modulate the system's thermodynamic response \cite{herrero2020thermodynamic,yan2022thermodynamic}.
While in a prior work the high efficiency of magic-angle TBG (MATBG) operating as a Quantum Otto Heat Engine (QOHE) \cite{singh2021magic} has been demonstrably verified, this present study extends the analysis considerably. We perform a comprehensive comparative investigation of MATBG under three major quantum thermodynamic cycles: the Quantum Otto Cycle (QOC), Quantum Carnot Cycle (QCC), and Quantum Stirling Cycle (QSC), and systematically examine its performance across various operational regimes, including heat engine, refrigerator, cold pump, and Joule pump modes, many of which were not previously explored. Additionally, we broaden the scope of the analysis by comparing MATBG to other graphene based systems, including monolayer graphene, AB-stacked bilayer graphene, and non-magic-angle twisted bilayer graphene, across all thermodynamic cycles and operating modes.

In section \ref{sec:TBG Theory}, we outline the theoretical framework describing the graphene systems under consideration, including monolayer graphene, bilayer graphene, and TBG. We present the corresponding low-energy Hamiltonians and examine how their energy spectra evolve under the influence of an external magnetic field. Section \ref{sec:Thermo framework} introduces the quantum thermodynamic framework used in this study, covering the general expressions for heat and work and detailed descriptions of the implemented thermodynamic cycles. In section \ref{sec:Result}, we present the results, identifying the operational regimes exhibited by different graphene platforms under various conditions. Section \ref{sec:analys} presents a comprehensive analysis of the results, where we compare the thermodynamic performance, such as efficiency, coefficient of performance (COP), and work exchange, across different cycles and operational regimes for various graphene systems. Finally, we summarize our findings and briefly discuss potential experimental realizations in section \ref{sec:conc}.

\section{Landau Levels of Twisted Bilayer Graphene}
\label{sec:TBG Theory}

Before introducing the model Hamiltonian for TBG, we briefly review the Landau level spectra of monolayer and bilayer graphene. We focus on their low-energy effective Hamiltonians near the Dirac points and examine how a perpendicular magnetic field modifies their energy levels. The monolayer graphene Hamiltonian  ~\citep{goerbig2011electronic,python2019}, and AB-Bernal stacked bilayer graphene Hamiltonian ~\citep{mccann2013electronic,mccann2006landau}  near the Dirac point in presence of magnetic field (\(B\)) is given by,






\begin{equation}
\small
\begin{aligned}
  h_{\boldsymbol{k}}^{\text{mono}}(B) &=  \frac{\xi\sqrt{2} \hbar v_f}{l_B}
  \begin{bmatrix}
  0 & \Pi^{\dagger} \\
  \Pi & 0
  \end{bmatrix}, \quad 
  h_{\boldsymbol{k}}^{\text{bi}}(B) = -\frac{\hbar^2}{m_{\text{eff}} l_B^2}
  \begin{bmatrix}
  0 & (\Pi^{\dagger})^2 \\
  (\Pi)^2 & 0
  \end{bmatrix}
\end{aligned}
\label{eq:mono_bi}
\end{equation}
where \( l_B = \sqrt{\hbar / eB} \) is the magnetic length, and \( m_{\text{eff}} \approx 0.035\,m_e \) is the effective electron mass in bilayer graphene~\citep{mccann2006landau}. \(\xi\) is the valley psuedo-spin, \(\Pi ,\Pi^{\dagger}\) are the ladder operators obeying \( [\Pi, \Pi^{\dagger}] = 1 \) \citep{python2019}. Solving Eq. \eqref{eq:mono_bi}, we obtain the Landau level spectra for both monolayer and bilayer, 


\begin{equation}
\small
\begin{aligned}
  E_n^{\text{mono}}(B) &= \pm \frac{\hbar v_f}{l_B} \sqrt{2n}, \quad
  E_n^{\text{bi}}(B) = \pm \hbar \omega_B \sqrt{n(n - 1)}, \quad n = 0, 1, 2, \dots
\end{aligned}
\label{eq:spec_mono_bi}
\end{equation}
where \(\omega_B=\frac{eB}{m_\text{eff}}\). Landau levels follow a $\sqrt{nB}$ scaling in monolayer graphene, with a four-fold degenerate zeroth level at zero energy~\citep{goerbig2011electronic}. In contrast, bilayer graphene, with $\sqrt{n(n-1)}B$ scaling, has both \(n=0 \) and \(n=1 \) at zero energy, yielding an eight-fold degenerate zeroth level.





The continuum model for twisted bilayer graphene at low energies consists of two Dirac-like Hamiltonians, which capture the intralayer electronic dynamics of each graphene layer, along with an interlayer coupling term that accounts for tunneling between the rotated layers~\citep{moire,python2019}. By restricting the continuum model to interactions within the first moiré shell, we obtain an effective eight-band Hamiltonian that describes the dominant electronic couplings in twisted bilayer graphene~\citep{moire,bistritzer2011moire,lopes2012continuum},

\begin{equation}
\small
    \mathcal{H}_{\mathbf{k},\theta}^{\text{TBG}}=\begin{bmatrix} h_{\mathbf{k}}^{\text{mono}}(\theta/2) & T_b  & T_{tr} & T_{tl} \\ T_b^{\dagger}   & h_{\mathbf{k_b}}^{\text{mono}}(-\theta/2)&0 &0  \\ T_{tr}^{\dagger} &0 & h_{\mathbf{k_{tr}}}^{\text{mono}}(-\theta/2)&0 \\ T_{tl}^{\dagger} &0 &0& h_{\mathbf{k_{tl}}}^{\text{mono}}(-\theta/2)
    \end{bmatrix}
    \label{eq:TBG}
\end{equation} 
In \eqref{eq:TBG}, $\mathbf{k}$ lies in the moiré Brillouin zone and $h_{\mathbf{k}}^{mono}(\theta)$ is the monolayer Dirac Hamiltonian rotated by \(\theta\) degrees. The vectors $\mathbf{k_j} = \mathbf{k} + \mathbf{q_j}$ for $\textbf{j} \in \{B, \text{tr}, \text{tl}\}$ describe momentum shifts due to interlayer tunneling near the Dirac points, are,
\begin{equation}
    \mathbf{q_b} = k_{\theta} \left( 0,-1 \right), \quad \mathbf{q_{tr}} = k_{\theta} \left( \frac{\sqrt{3}}{2},\frac{1}{2} \right), \quad \mathbf{q_{tl}} = k_{\theta} \left( -\frac{\sqrt{3}}{2},\frac{1}{2} \right)
    \label{eq : Hopping vectors}
\end{equation}
where \( k_{\theta} = \frac{8\pi}{3a} \sin\left( \frac{\theta}{2} \right) \), with \( a = 2.46\,\text{\AA} \) denoting the graphene lattice constant. The interlayer hopping matrices are,
\begin{equation}
    T_b = \omega \begin{bmatrix} 1 & 1 \\ 1 & 1 \end{bmatrix}, \quad T_{tr} = \omega \begin{bmatrix} e^{-i \phi} & 1 \\ e^{i \phi} & e^{-i \phi} \end{bmatrix}, \quad T_{tl} = \omega \begin{bmatrix} e^{i \phi} & 1 \\ e^{-i \phi} & e^{i \phi} \end{bmatrix}
    \label{eq:Hoppingmatrix}
\end{equation}

\begin{figure}[!htbp]
    \centering
    \includegraphics[width=1\linewidth]{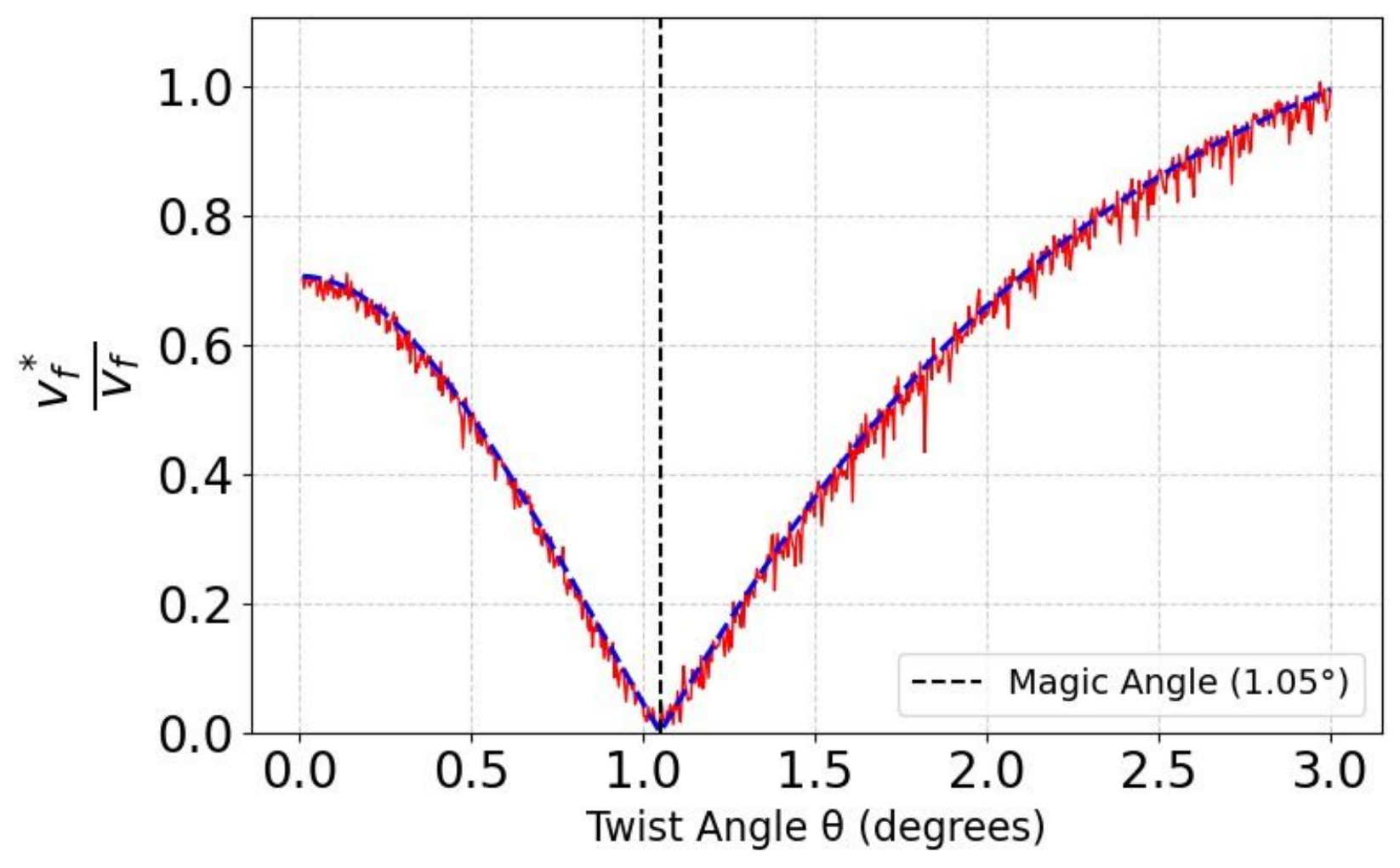}
    \caption{Renormalized Fermi velocity vs twist angle $\theta$,  exact 8-band Hamiltonian results (solid red line) with the approximated 2-band model (dashed blue line). The normalized velocity tends to zero in both cases at the magic-angle (\(\theta=1.05^o\))}
    \label{fig1}
\end{figure}

where $\omega = 0.110 \ \text{eV}$ is the interlayer hopping energy, and $\phi = \frac{2\pi}{3}$.The Hamiltonian $\mathcal{H}_{\mathbf{k},\theta}^{TBG}$ can be simplified by expanding around the Dirac point of the first graphene layer. By separating momentum-dependent and independent terms and expanding the momentum-dependent terms to leading order in $\mathbf{k}$, the independent terms have zero eigenvalues. As a result, only the momentum-dependent part contributes, leading to the low-energy Hamiltonian near the Dirac cone \citep{python2019,moire},

\vspace{-1.7em}
{\small
\begin{equation}
    \mathcal{H}_{\mathbf{k \sim 0},\theta}^{TBG} = \frac{\hbar v_f}{1 + 6\alpha_{\theta}^2} \left( \mathbf{\sigma^* \cdot k} - \sum_j T_j \left( h^{mono}_{\mathbf{q_j}} \right)^{-1^\dagger} \left( \mathbf{\sigma^* \cdot k} \right) \left( h^{mono}_{\mathbf{q_j}} \right)^{-1} T_j^{\dagger} \right)
    \label{eq:TBG_knear0}
\end{equation}
}
\vspace{-1.7em}

The first term in Eq.\eqref{eq:TBG_knear0} represents the isolated Dirac cone of a single graphene layer, while the second term captures the contributions from neighboring Dirac cones in the adjacent layer. Eq.\eqref{eq:TBG_knear0} can be further simplified by substituting (\ref{eq : Hopping vectors}) and (\ref{eq:Hoppingmatrix}) for the second term to obtain the reduced two-band model for TBG \citep{python2019},

\vspace{-1.5em}
\begin{equation}
    \mathcal{H}_{\mathbf{k \sim 0},\theta}^{TBG} = \hbar v_f^* \mathbf{\sigma^* \cdot k} \quad, \quad v_f^* = v_f \frac{1 - 3 \alpha_{\theta}^2}{1 + 6 \alpha_{\theta}^2}
    \label{eq:TBG_2band}
\end{equation}
\vspace{-1.5em}

Here, \( v_f^* \) is the renormalized Fermi velocity, and \( \alpha_{\theta} = \frac{\omega}{\hbar v_f k_{\theta}} \) defines the dimensionless interlayer coupling parameter. Notably, \( v_f^* \) vanishes at the so-called magic angle \( \theta^* = 1.05^\circ \), as depicted in Fig.~\ref{fig1}. where the electronic band structure exhibits flat bands. By including the effects of a perpendicular magnetic field within the continuum model (see, Eqs.~\eqref{eq:mono_bi},~\eqref{eq:TBG}), one can numerically calculate the resulting Landau level spectrum for twisted bilayer graphene~\citep{python2019,zhang2019landau,moon2012energy}.

\section{Quantum Thermodynamic Cycles}
\label{sec:Thermo framework}

\begin{figure}[!htbp]
    \centering
    \includegraphics[width=1\linewidth]{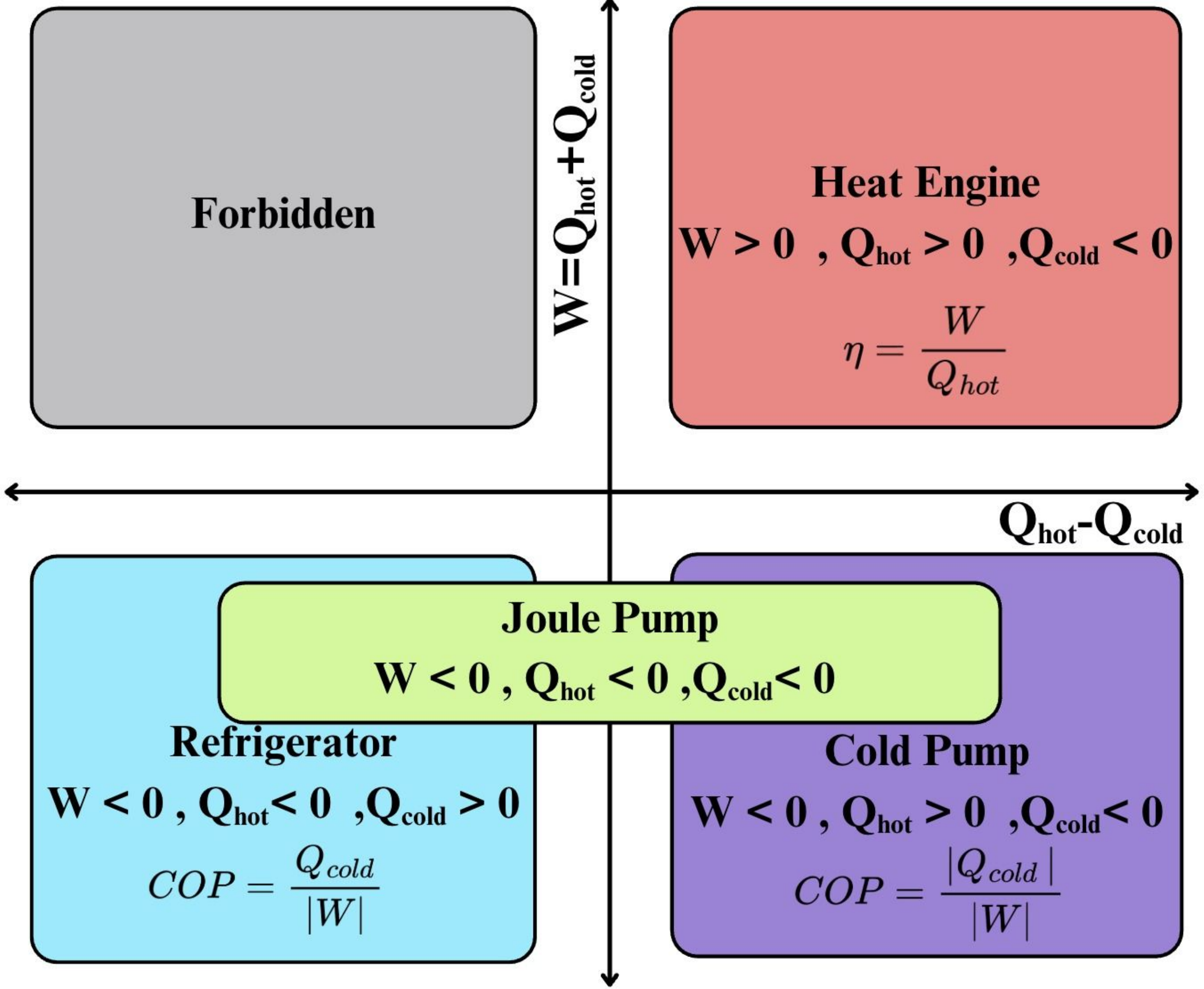}
    \caption{Operational regimes in any quantum thermodynamic cycle. The heat engine is present in the first quadrant, the refrigerator in the third quadrant, and the cold pump in the fourth quadrant. The Joule pump can appear in the third or fourth quadrant, depending on the relative magnitudes of heat exchanges. Any other heat and work relation violates the first or second law of thermodynamics (see, Appendix A).}
    \label{fig2}
\end{figure}

Quantum thermodynamics generalizes classical thermodynamics to microscopic systems where quantum effects are significant. It describes the functioning of quantum heat engines, refrigerators, cold pumps, and Joule pumps in terms of energy and entropy exchange processes~\citep{prakash2022impurity,scharf2020topological,quan2009quantum}. The operational regime of a device is determined by the signs of heat flows (\(Q_{\text{hot}}, Q_{\text{cold}}\)) and work (\(\text{W}\)). Fig. ~\ref{fig2} summarizes all possible regimes and their corresponding performance metrics. Here, \(\text{W} > 0\) indicates work done by the system, \(\text{W} < 0\) represents work done on the system, while \(Q > 0\) denotes heat absorbed, and \(Q < 0\) heat released by the system \citep{quan2009quantum,lucio2025innovative}. For a system with Hamiltonian \( H \), eigenstates \( \{ |n\rangle \} \), and corresponding eigenvalues \( \{ E_n \} \), the thermal state is given by,

\vspace{-0.8em}
\begin{equation}
\rho = \sum_n p_n |n\rangle\langle n| \quad, \quad p_n= \frac{e^{-\beta E_n}}{Z}\quad, \quad Z = \sum_n e^{-\beta E_n}
\label{eq: Thermal_state}
\end{equation}
\vspace{-1.0em}

\noindent where, \(p_n\) are the occupation probability of the state \(\ket{n}\), and \( \beta = \frac{1}{k_B T} \) the inverse temperature of the reservoir. The entropy of a quantum state \( \rho \) is defined as,

\vspace{-1em}
\begin{equation}
S = -k_B \text{Tr}(\rho \ln \rho),
\label{eq:von_neumann_entropy}
\end{equation}
\vspace{-2.0em}

\noindent with internal energy given by,

\vspace{-2em}
\begin{equation}
U = \langle H \rangle = \sum_n p_n E_n.
\label{eq:internal_energy}
\end{equation}
\vspace{-1.8em}

During an infinitesimal transformation, the first law of thermodynamics is expressed as,

\vspace{-1.5em}
\begin{equation}
dU = \delta Q - \delta \text{W},\ \  \text{with} \ \ \ \delta Q = \sum_n E_n \, dp_n, \quad \delta \text{W} =- \sum_n p_n \, dE_n.
\label{eq: Heat_work}
\end{equation}
\vspace{-1em}

On completion of a cycle, $dU=0$, implying the total work exchanged equals the sum of heat transfers. In order to define the performance characteristics of the thermodynamic cycles, we use the following metrics ~\citep{Quan2007,lucio2025innovative,landi2021irreversible}.

\begin{equation}
\begin{aligned}
    &\bullet\ \text{Heat Engine:}\\ & \quad \quad \eta = \frac{W}{Q_{\text{hot}}}, \quad \quad\text{Coefficient of merit} = W\times \frac{\eta}{\eta_c}, \\
    &\bullet\ \text{Refrigerator:} \\ & \quad \quad \text{COP}_{\text{R}} = \frac{Q_{\text{cold}}}{|W|},\quad \quad\text{Coefficient of merit} = Q_{\text{cold}}\times \frac{\text{COP}_\text{R}}{\text{COP}_c}, \\
    &\bullet\ \text{Cold Pump:} \\ &\quad \quad \text{COP}_{\text{CP}} = \frac{|Q_{\text{cold}}|}{|W|},\quad \quad\text{Coefficient of merit} = |Q_{\text{cold}}|\times \text{COP}_\text{CP}, \\
    & \bullet\ \text{Joule Pump:} \\ &\quad \quad \Delta S=-(\frac{Q_{\text{hot}}}{T_h}+\frac{Q_{\text{cold}}}{T_c}),
\end{aligned}
\label{eq:performance_metrics}
\end{equation}

Where \(\eta_c\) and \(\text{COP}_c\) are the Carnot efficiency and coefficient of performance, respectively. For heat engines, refrigeration, and cold pumps, the performance metric is chosen such that the numerator represents the target energy transfer, and the denominator corresponds to the cost required for the system’s operation. It is important to note that the Joule pump operates by converting the supplied work entirely into heat distributed between both reservoirs. As a result, its coefficient of performance is given by \( \text{COP}_\text{JP} = \frac{|Q_\text{hot}| + |Q_\text{cold}|}{|W|} = 1 \), irrespective of the system~\citep{scharf2020topological}, hence we analyze the entropy produced per cycle in the Joule pump regime. A smaller value of \(\Delta S\) indicates a more reversible operation, signifying reduced entropy generation and minimal disturbance to the thermal reservoirs. 

In the following section, we find the expressions for the heat transfers in the thermalization stroke, work exchange, and the performance of QCC, QOC, and QSC.

\subsection{Quantum Otto Cycle}
\begin{figure}[!htbp]
    \centering
    \includegraphics[width=0.95\linewidth]{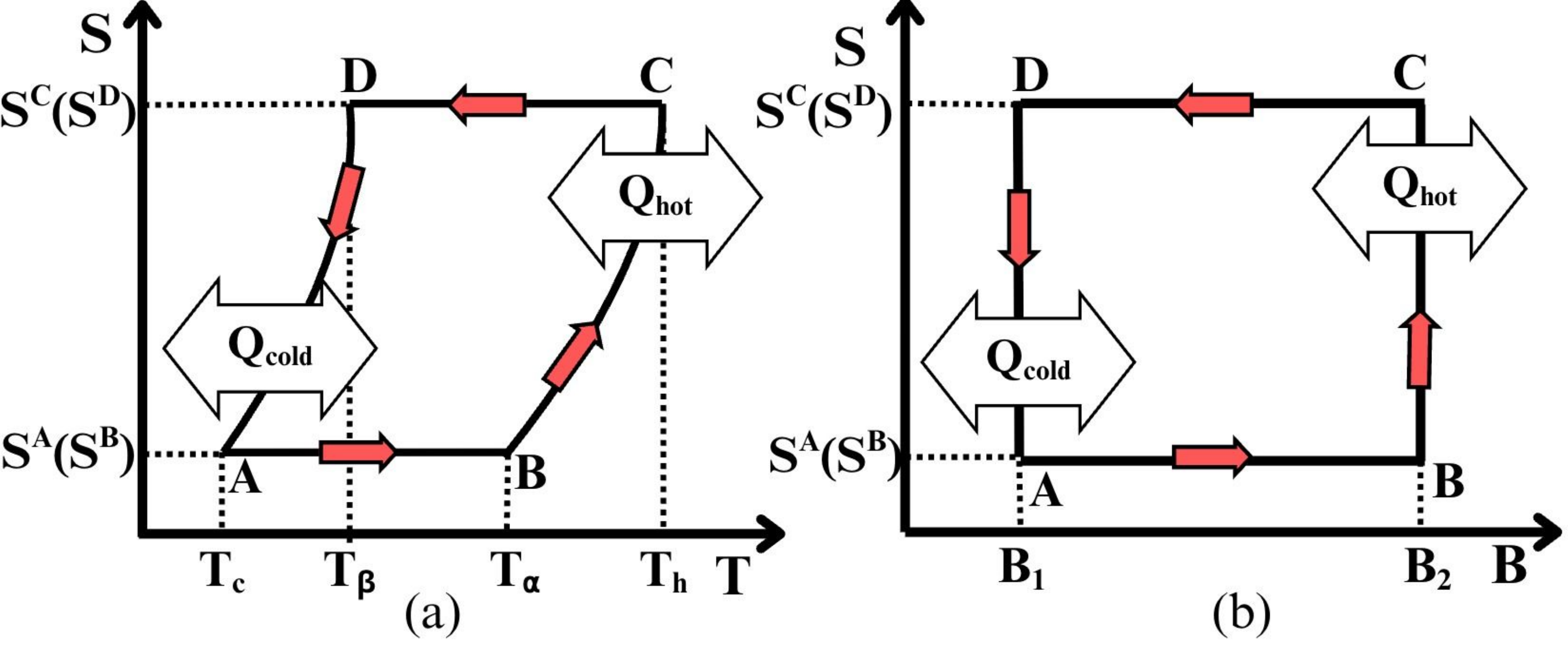}
    \caption{(a) Entropy–temperature (S–T) and (b) entropy–magnetic field (S–B) diagrams for the QOC. Strokes \( \text{A} \rightarrow \text{B} \) and \( \text{C} \rightarrow \text{D} \) are adiabatic, while \( \text{B} \rightarrow \text{C} \) and \( \text{D} \rightarrow \text{A} \) are isochoric.}
    \label{fig3}
\end{figure}


{
 The QOC can be implemented in two ways: under the general or the strict adiabatic condition. The general condition requires the system’s entropy to remain constant at the end of the adiabatic stroke, which is what thermodynamic adiabatic condition entails. The strict condition demands that the occupation probabilities of the instantaneous eigenstates remain unchanged, obeying to the quantum adiabatic theorem \cite{Quan2007}. The strict condition can be realized by slowly varying magnetic field so as to avoid energy level transitions. While a strict adiabatic process necessarily satisfies the general condition (see Eq. \eqref{eq:von_neumann_entropy}), the converse is not true, as the general condition can be enforced through mechanisms that redistribute populations while conserving entropy \citep{Quan2007,quan2009quantum,pena2020otto}. A general adiabatic condition only implies the strict if the energy levels scale by same ratio }(see, Appendix C). In such cases, the final state remains thermal, and both conditions yield identical results. The four stages of the QOC are illustrated in Fig.~\ref{fig3}.

\begin{itemize}
    \item \textbf{(A$\rightarrow$B) Adiabatic Stroke}: The system, initially at temperature \( T_\text{c} \) and magnetic field \( B_1 \), undergoes an adiabatic transformation where the magnetic field is changed to \( B_2 \). Under the general condition, the system evolves to a new effective temperature \( T_\alpha \) to conserve entropy. Under the strict condition, the final state remains thermal only if all energy gaps scale uniformly \citep{quan2009quantum},
    \vspace{-0.5em}
    \begin{equation}
    \begin{split}
        &S^{\text{A}}(B_1,T_\text{c}) = S^{B}(B_2,T_\alpha) \quad \text{(General)}, \\
        &p_n^{\text{A}}(B_1,T_\text{c}) = p_n^{B}(B_2) \quad \forall n \quad \text{(Strict)}
    \end{split}
    \label{eq:ottoaida1}
    \end{equation}
    \vspace{-1em}
    
    We can use the above expression for the general case to determine the effective temperature \(T_\alpha\) to construct the thermal state \(p_n^{B}(B_2,T_\alpha)\).
    
    \item \textbf{(B$\rightarrow$C) Isochoric stroke}: At fixed magnetic field \( B_2 \), the system is brought in contact with a hot thermal reservoir, allowing it to thermalize to \( T_\text{h} \). System exchanges heat, and the occupation probabilities adjust to reflect a thermal state at \( T_\text{h} \) and \( B_2 \),

    \vspace{-1.0em}
    \begin{equation}
    \begin{split}
    &Q_{\text{hot}}^{\text{gen}} = \sum_n E_n(B_2)\left[p_n^{\text{C}}(B_2,T_\text{h}) - p_n^{B}(B_2,T_\alpha)\right] \quad \text{(General)}, \\ 
    & Q_{\text{hot}}^{\text{str}} = \sum_n E_n(B_2)\left[p_n^{\text{C}}(B_2,T_\text{h}) - p_n^{\text{A}}(B_1,T_\text{c})\right] \quad \text{(Strict)}
    \label{eq:otto heat1}
    \end{split}
    \end{equation}
    \vspace{-1.4em}
    
    \item \textbf{(C$\rightarrow$D) Adiabatic stroke}: The magnetic field is decreased from \( B_2 \) to \( B_1 \) without the exchange of heat. For the general case, the system reaches an effective temperature of \(T_\beta\). For a strict case, the final state is only thermal if all energy gaps scale uniformly \citep{quan2009quantum},

    \vspace{-1.4em}
    \begin{equation}
    \begin{split}
        &S^{\text{C}}(B_2,T_\text{h}) = S^{\text{D}}(B_1,T_\beta)\ \quad \text{(General)}, \\ &p_n^{\text{C}}(B_2,T_\text{h}) = p_n^{\text{D}}(B_1)\quad \forall n\ \quad \text{(Strict)}
    \end{split}
        \label{eq:ottoaida2}
    \end{equation}
    \vspace{-1em}
    
     For the general condition, we can determine the effective temperature \(T_\beta\) from the above expression, which then allows us to construct the thermal state \(p_n^{\text{D}}(B_2, T_\beta)\).

    \item \textbf{(D$\rightarrow$A) Isochoric stroke}: At fixed magnetic field \( B_1 \), the system is kept in contact with the cold reservoir to thermalize to initial temperature by exchanging heat. 

    \vspace{-1.4em}
    \begin{equation}
    \begin{split}
         & Q^{\text{gen}}_{\text{cold}} = \sum_n E_n(B_1)\left[p_n^{\text{A}}(B_1,T_\text{c}) - p_n^{\text{D}}(B_1,T_\beta)\right] \quad (\text{General}), \\ &Q^{\text{str}}_{\text{cold}} = \sum_n E_n(B_1)\left[p_n^{\text{A}}(B_1,T_\text{c}) - p_n^{\text{C}}(B_2,T_\text{h})\right] \quad (\text{Strict})
         \label{eq:otto heat2}
    \end{split}
    \end{equation}
    \vspace{-2.2em}

\end{itemize}

The net work output is $\text{W}^{\text{str}} = Q_{\text{hot}}^{\text{str}} + Q_{\text{cold}}^{\text{str}}$ for strict case and $\text{W}^{\text{gen}} = Q_{\text{hot}}^{\text{gen}} + Q_{\text{cold}}^{\text{gen}}$ for general case. We can then calculate the performance for each operational phase using Eq.\eqref{eq:performance_metrics}.

\subsection{Quantum Carnot Cycle}
\begin{figure}[!htbp]
    \centering
    \includegraphics[width=0.95\linewidth]{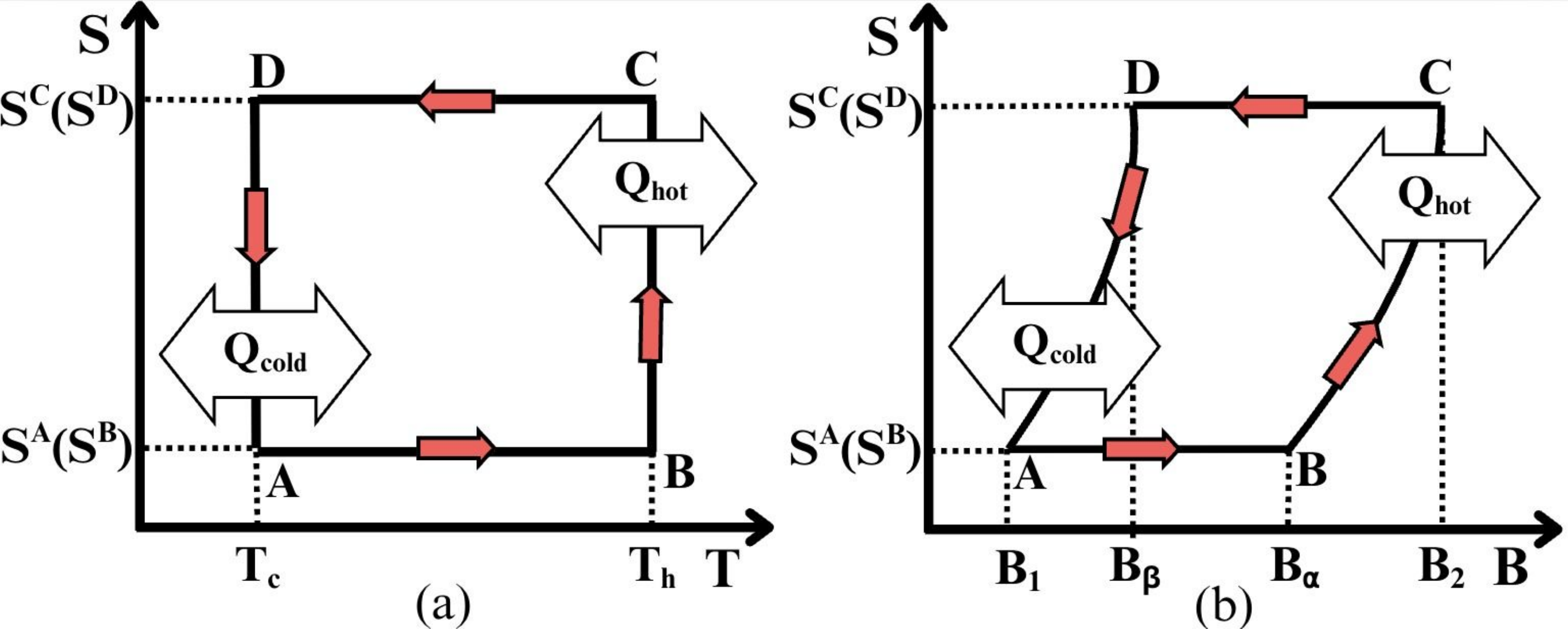}
    \caption{(a) Entropy–temperature (S–T) and (b) entropy–magnetic field (S–B) diagrams for the QCC. Strokes \( \text{A} \rightarrow \text{B} \) and \( \text{C} \rightarrow \text{D} \) are adiabatic, while \( \text{B} \rightarrow \text{C} \) and \( \text{D} \rightarrow \text{A} \) are isothermal.}
    \label{fig4}
\end{figure}

The QCC for a graphene system with a tunable magnetic field ($B$) is illustrated in the diagram in Fig.~\ref{fig4}. It consists of two adiabatic and two isothermal strokes, with heat exchange occurring only during the isothermal processes. { To construct a  reversible cycle, the energy levels must satisfy the scaling condition mentioned in ~\cite{quan2009quantum}, which requires the system to be in thermal state at the end of the strict adiabatic process. This condition is satisfied by monolayer and bilayer graphene but not by twisted bilayer graphene (see, Appendix C)}. Therefore, in this section, we implement the reversible QCC under the general adiabatic condition, which does not put any constraint on the energy eigenvalues of the system. The four stages of the QCC are as follows:

\begin{itemize}
    \item \textbf{(A$\rightarrow$B) Adiabatic stroke}: 
    The system, initially at temperature \( T_\text{c} \) and magnetic field \( B_1 \), undergoes an adiabatic transformation to magnetic field \( B_\alpha \), during which its temperature increases to \( T_\text{h} \) with no entropy change.
    \vspace{-0.5em}
    \begin{equation}
    S^{\text{A}}(B_1,T_\text{c}) = S^{B}(B_\alpha,T_\text{h})
        \label{eq:carnotaida1}
    \end{equation}
    \vspace{-2.5em}

    \item \textbf{(B$\rightarrow$C) Isothermal stroke}:
    The system undergoes an isothermal transformation at temperature \( T_\text{h} \), where the magnetic field changes from \( B_\alpha \) to \( B_2 \), allowing heat exchange with the hot reservoir while keeping the temperature fixed. From Eq. (\ref{eq: Heat_work})

    \vspace{-1.0em}
    \begin{equation}
        Q_{\text{hot}}=\sum_n \int_{B_\alpha}^{B_2} E_n(B)\frac{\partial p_n(B,T_\text{h})}{\partial B}dB
        \label{eq: isothermal heat}
    \end{equation}
    \vspace{-1.0em}

    Eq. (\ref{eq: isothermal heat}) is reduced (see, Appendix D) to, 

    \vspace{-1.5em}
   \begin{equation}
    \begin{aligned}
    Q_{\text{hot}} &= T_\text{h}\left[S^{\text{C}}(B_2,T_\text{h}) - S^{B}(B_\alpha,T_\text{h})\right] \\
                   &= T_\text{h}\left[S^{\text{C}}(B_2,T_\text{h}) - S^{\text{A}}(B_1,T_\text{c})\right]
    \end{aligned}
    \label{eq:Qhot}
    \end{equation}
    \vspace{-1.8em}

    \item \textbf{(C$\rightarrow$D) Adiabatic stroke}:
    The system undergoes an adiabatic transformation, where the magnetic field is changed from \( B_2 \) to \( B_\beta \), and the temperature reduces from \( T_\text{h} \) to \( T_\text{c} \), with no heat exchange, and entropy is conserved, thus

    \vspace{-2.5em}
    \begin{equation}
    S^{\text{C}}(B_2,T_\text{h}) = S^{\text{D}}(B_\beta,T_\text{c})
    \label{eq:carnotaida2}
    \end{equation}
    \vspace{-2.0em}

    \item \textbf{(D$\rightarrow$A) Isothermal stroke}:
    The system undergoes an isothermal process at temperature \( T_\text{c} \) connected to a cold reservoir, with the magnetic field changed from \( B_\beta \) to \( B_1 \), bringing the system back to its initial state while absorbing or releasing heat to the cold reservoir.

    \vspace{-2.0em}
    \begin{equation}
\begin{aligned}
    Q_{\text{cold}} &= T_\text{c}\left[S^{\text{A}}(B_1,T_\text{c}) - S^{\text{D}}(B_\beta,T_\text{c})\right] \\
                           &= T_\text{c}\left[S^{\text{A}}(B_1,T_\text{c}) - S^{\text{C}}(B_2,T_\text{h})\right]
\end{aligned}
\label{eq:Qcold}
\end{equation}
\vspace{-2.0em}

\end{itemize}

The net work per cycle is $
    \text{W} = Q_{\text{hot}} + Q_{\text{cold}} = (T_\text{h} - T_\text{c})[S_C(B_2,T_\text{h}) - S_A(B_1,T_\text{c})]$. We can calculate the performance of the observed operational regime using Eq. \eqref{eq:performance_metrics}. Since the QCC is implemented under reversible conditions, it operates at Carnot efficiency, \(\eta_C = 1 - \frac{T_\text{c}}{T_\text{h}}\) when functioning as a heat engine, and achieves the Carnot coefficient of performance, \(\text{COP}_c = \frac{T_\text{c}}{T_\text{h} - T_\text{c}}\) when operating as a refrigerator.


\subsection{Quantum Stirling Cycle}
\begin{figure}[!htbp]
    \centering
    \includegraphics[width=0.95\linewidth]{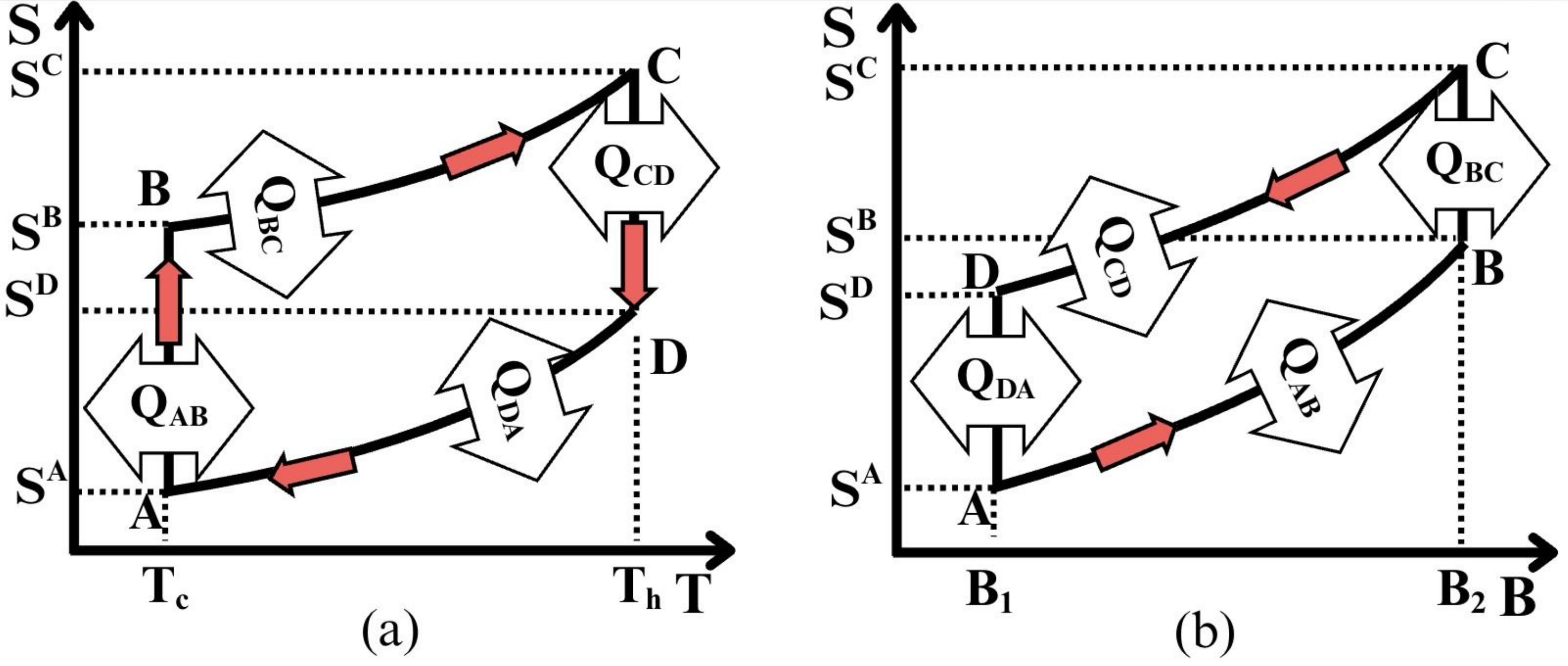}
    \caption{(a) Entropy–temperature (S–T) and (b) entropy–magnetic field (S–B) diagrams for the QSC. Strokes \( \text{A} \rightarrow \text{B} \) and \( \text{C} \rightarrow \text{D} \) are isothermal, while \( \text{B} \rightarrow \text{C} \) and \( \text{D} \rightarrow \text{A} \) are isochoric.}
    \label{fig5}
\end{figure}

\begin{figure*}[t]
    \centering
    \begin{minipage}[t]{0.45\linewidth}
        \centering
        \includegraphics[width=\linewidth]{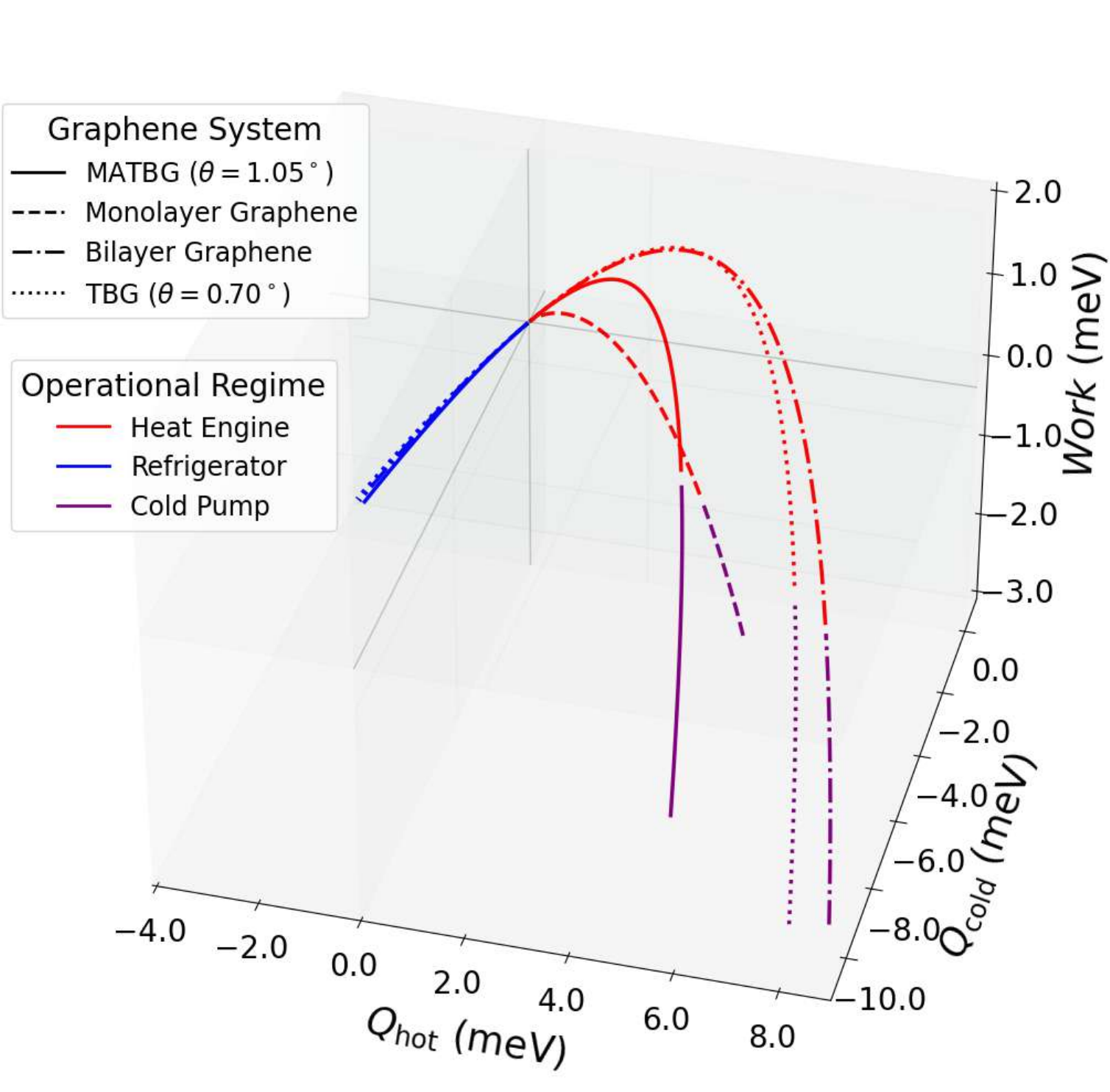}
        \vspace{2pt}
        
        {\small (a) General adiabatic condition}
    \end{minipage}
    \hfill
    \begin{minipage}[t]{0.45\linewidth}
        \centering
        \includegraphics[width=\linewidth]{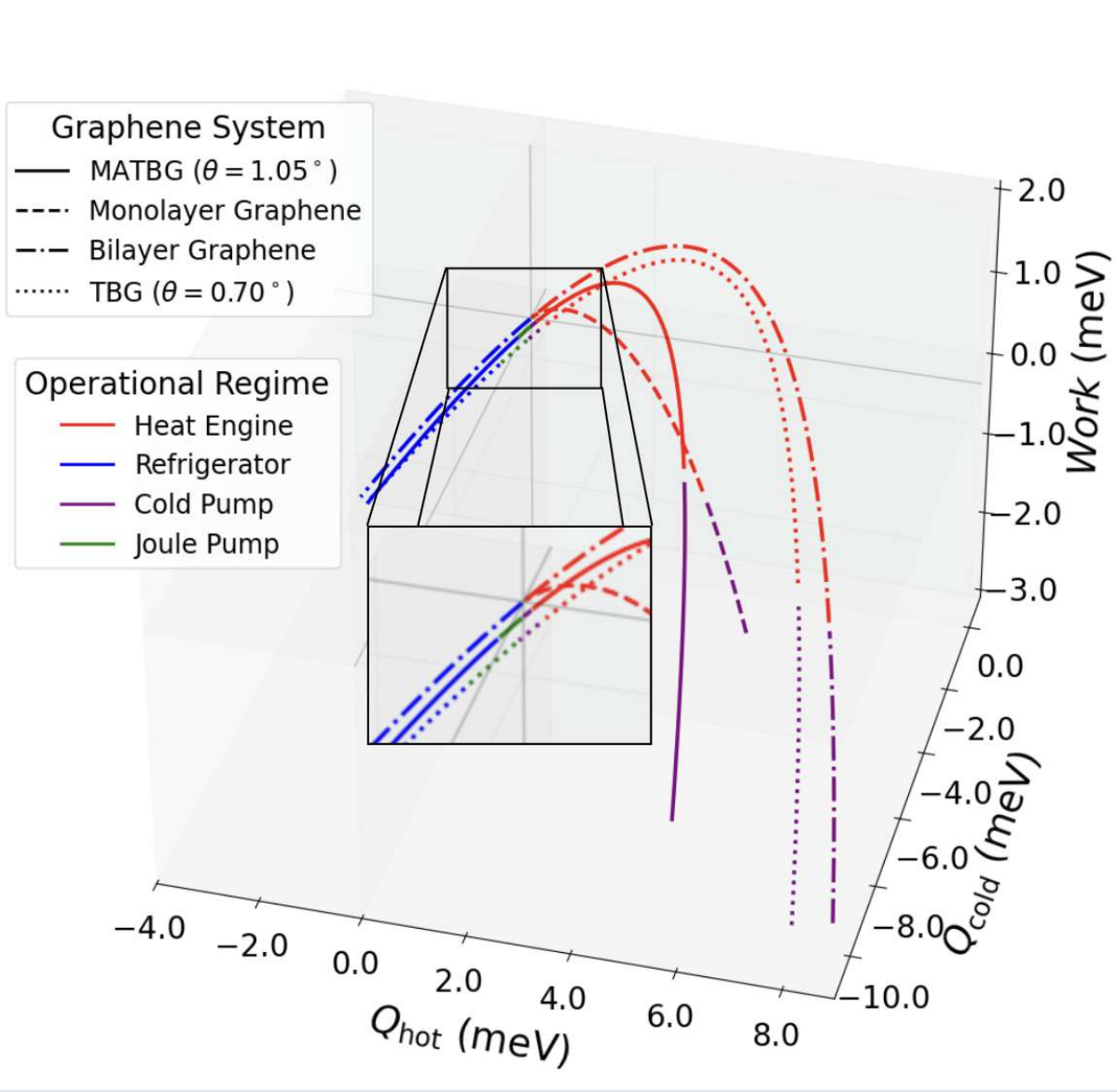}
        \vspace{2pt}
        
        {\small (b) Strict adiabatic condition}
    \end{minipage}
    
    \caption{Operational phases of the QOC under (a) general adiabatic condition and (b) strict adiabatic condition, for different graphene systems. Parameters : $T_\text{h}=150$ K, $T_\text{c}=50$ K, and $B_1=1$ T, with $r_c$ varied from $0.5$ to $4$.}
    \label{fig6}
\end{figure*}

The Stirling cycle consists of two isothermal and two isochoric strokes as shown in Fig. (\ref{fig5}).
\begin{itemize}
    \item \textbf{(A$\rightarrow$B) Isothermal stroke}: The system starts at magnetic field \( B_1 \) and temperature \( T_\text{c} \), connected to cold reservoir, undergoes an isothermal process as \( B \) is slowly varied to \( B_2 \). It remains equilibrated throughout, and the heat exchanged \( Q_{AB} \) is,

    \vspace{-2.0em}
    \begin{equation}
        Q_{\text{AB}}= T_\text{c}[S^{B}(B_2,T_\text{c}) - S^{\text{A}}(B_1,T_\text{c})]
        \label{eq:sterlingheat1}
    \end{equation}
    \vspace{-2.5em}

    \item \textbf{(B$\rightarrow$C) Isochoric stroke}: In this stroke, the magnetic field is fixed at \( B_2 \), and the system is connected to the hot reservoir, raising its temperature from \( T_\text{c} \) to \( T_\text{h} \). The heat exchanged, \( Q_{BC} \), is,
    \vspace{-0.7em}
         \begin{equation}
        Q_{\text{BC}}=  \sum_n E_n(B_2)\left[p_n^{\text{C}}(B_2,T_\text{h}) - p_n^{B}(B_2,T_\text{c})\right]
        \label{eq:sterlingheat2}
    \end{equation}  
    \vspace{-1.5em}

    \item \textbf{(C$\rightarrow$D) Isothermal stroke}: In this stroke, the magnetic field is varied from \( B_2 \) to \( B_1 \) maintained the system at a constant temperature \( T_\text{h} \). Both heat exchange and work are involved in this process, with heat \( Q_{CD} \) exchanged with the hot reservoir,

    \vspace{-1.3em}
    \begin{equation}
        Q_{\text{CD}}=  T_\text{h}[S^{\text{D}}(B_1,T_\text{h}) - S^{\text{C}}(B_2,T_\text{h})]
        \label{eq:sterlingheat3}
    \end{equation}  

    \vspace{-1.0em}

    \item \textbf{(D$\rightarrow$A) Isochoric stroke}: In the final stroke, the magnetic field remains fixed at \( B_1 \), while the system cools from \( T_\text{h} \) to its initial temperature \( T_\text{c} \). During this thermalization, it releases heat \( Q_{DA} \) to the cold reservoir, thereby completing the cycle and restoring the system to its initial state,

    \begin{equation}
        Q_{\text{DA}}=  \sum_n E_n(B_1)\left[p_n^{\text{A}}(B_1,T_\text{c}) - p_n^{\text{D}}(B_1,T_\text{h})\right]
        \label{eq:sterlingheat4}
    \end{equation} 
    \vspace{-1.5em}
\end{itemize}

We then have, $Q_{\text{hot}}=Q_{\text{BC}}+Q_{\text{CD}},Q_{\text{cold}}=Q_{\text{DA}}+Q_{\text{AB}}$ and $\text{W}=Q_{\text{hot}}+Q_{\text{cold}}$. The performance of any operational regime can be calculated using Eq. \eqref{eq:performance_metrics}.


\section{Results}
\label{sec:Result}

In this section, we consider hot reservoir temperature $T_\text{h} = 150\,\text{K}$ and cold reservoir temperature $T_\text{c} = 50\,\text{K}$. The tunable external magnetic fields used to drive the thermodynamic cycles are \( B_1 \) and \( B_2 \). We fix \( B_1 \) at \( 1\, T \), and vary the compression ratio ($r_c = B_2/B_1$) to investigate how the performance metrics of the cycle, such as efficiency, COP, work done, etc. For computational feasibility and consistency across systems, we truncate the Landau level spectrum to include the lowest 500 levels, which yields results with reasonable accuracy while capturing the essential physics of the thermodynamic behavior \cite{python2019,singh2021magic}. The numerical codes used to generate the results presented in this work are available in \citep{MATBGCode}.

\subsection{Operational phases in graphene systems}

Fig.~\ref{fig6} shows different operational regimes present during QOC using general and strict adiabatic conditions. We start from \(r_c=0.5\) and increase it to observe different operational phases. During QOC under general adiabatic conditions (Fig.~\ref{fig6}(a)), all systems are in cold pump phase at compression ratios (\(r_c < 1\)). As \(r_c\) increases, they transition into the heat engine regime. For monolayer and bilayer graphene, both adiabatic conditions lead to the same outcome, with monolayer graphene remaining in the heat engine regime with increasing \(r_c\), and bilayer graphene eventually shifting to the refrigeration regime. TBG under the general adiabatic condition follows a similar pattern to bilayer, transitioning from cold pump to heat engine, and finally to refrigeration phase.

\begin{figure}[!htbp]
    \centering
    \includegraphics[width=0.95\linewidth]{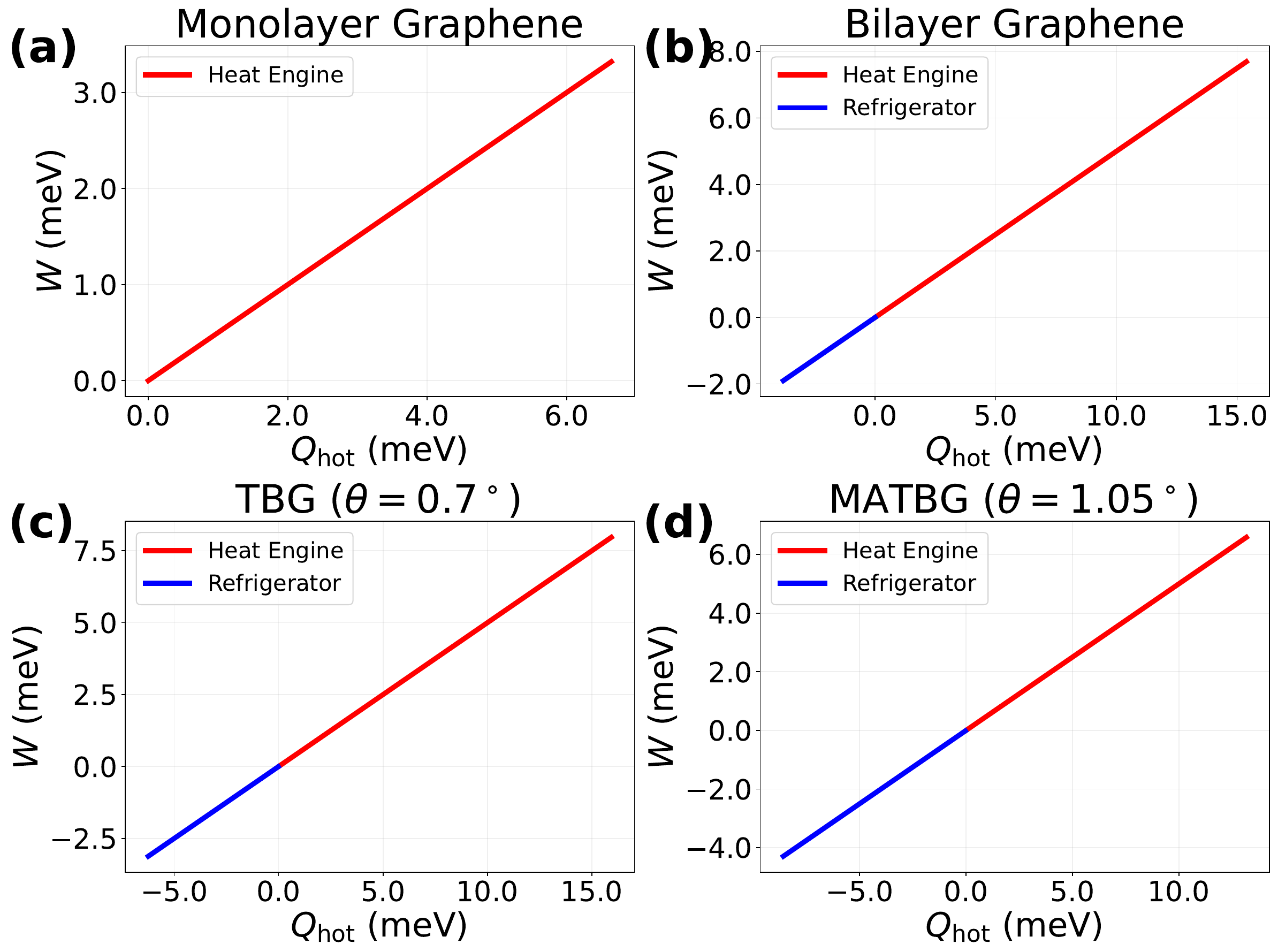}
    \caption{Operational phases of the QCC for different graphene systems. Parameters : $T_\text{h}=150$ K, $T_\text{c}=50$ K, and $B_1=1$ T, with $r_c$ varied from $0.5$ to $4$.}
    \label{fig7}
\end{figure}

However, under strict adiabatic conditions (Fig. \ref{fig6} (b)), TBG at both the magic angle (\(\theta = 1.05^\circ\)) and non-magic angle (\(\theta = 0.7^\circ\)) exhibits additional features. As \(r_c\) increases, these systems, initially in the cold pump phase, move to a heat engine phase, then to a second cold pump phase, then to a Joule pump phase, and finally to a refrigeration phase. These intermediate regimes appear only within a narrow window of compression ratios and are absent under the general adiabatic condition. The work output in the second cold pump phase and the Joule pump regime is relatively small.

Fig.~\ref{fig7} shows the operational regimes of graphene systems under the reversible QCC with increasing \(r_c\). Only heat engine and refrigeration phases are observed, with monolayer graphene exhibiting solely the heat engine phase; other systems transition from heat engine to refrigeration phase with increasing \(r_c\). Since reversible QCC should have zero entropy change for the universe, we must have,

\begin{equation}
    \Delta S_{\text{uni}}=\frac{Q_{\text{hot}}}{T_\text{h}}+\frac{Q_{\text{cold}}}{T_\text{c}}=0 \implies \frac{Q_{\text{hot}}}{Q_{\text{cold}}}\le 0 \ \ \text{and}\ \\  \frac{|Q_{\text{hot}}|}{|Q_{\text{cold}}|}=\frac{T_\text{h}}{T_\text{c}} \ge1 
    \label{eq: QCC condition}
\end{equation}

As a result, neither the cold pump phase, characterized by \( |Q_{\text{cold}}| > |Q_{\text{hot}}| \) nor the Joule pump phase, where both \( Q_{\text{cold}} < 0 \) and \( Q_{\text{hot}} < 0 \) is observed for reversible QCC. Eq. ~\eqref{eq: QCC condition} implies that \(Q_\text{hot} = -\frac{T_\text{h}}{T_\text{c}} Q_\text{cold}\), which results in all plots in Fig.~\ref{fig7} to have the same slope.

\begin{figure}[!htbp]
    \centering
    \includegraphics[width=0.95\linewidth]{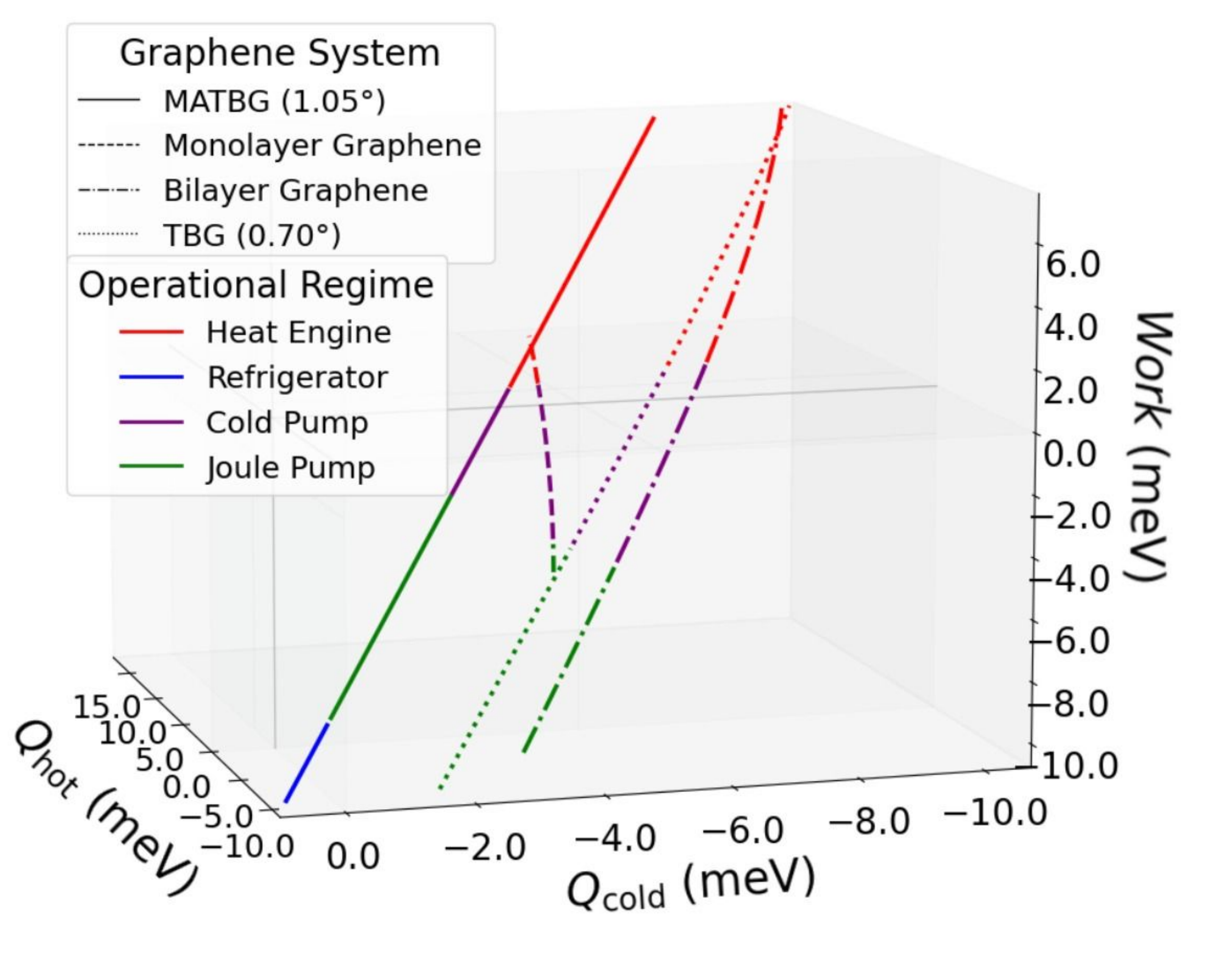}
    \caption{Operational phases of the QSC for different graphene systems. Parameters : $T_\text{h}=150$ K, $T_\text{c}=50$ K, and $B_1=1$ T, with $r_c$ varied from $0.3$ to $4$.}
    \label{fig8}
\end{figure}

For the QSC (Fig.~\ref{fig8}), we observe heat engine, cold pump, and Joule pump phases across all graphene systems. For the temperatures of reservoirs taken, we notice a refrigeration phase emerging only in the MATBG at low \(r_c\). In all systems except MATBG, increasing the \(r_c\) drives a transition from the Joule pump phase to the cold pump phase, and eventually to the heat engine phase at higher compression ratios. MATBG begins in the refrigeration phase, then follows a similar transition path as the other systems. Unlike strict QOC, we observe finite work input for the Joule pump phases.

\begin{figure}[!htbp]
    \centering
    \includegraphics[width=0.95\linewidth]{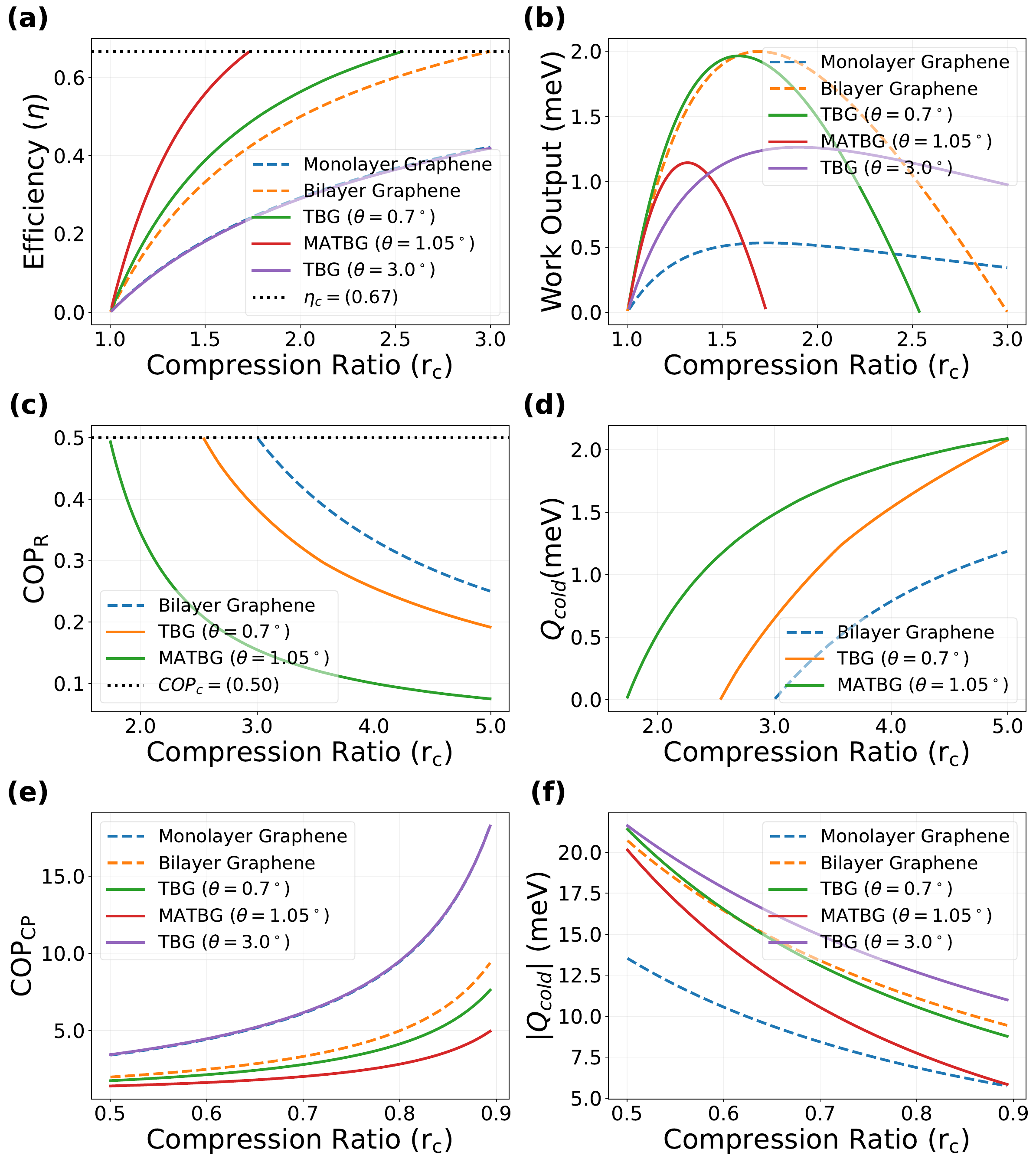}
    \caption{Performance of the QOC under general adiabatic conditions as a function of the compression ratio \( r_c \) for different graphene systems. Subplots show (a) efficiency and (b) work output in the heat engine regime, (c) COP and (d) Refrigeration Output in the refrigerator regime, and (e) COP and (f) heat output in the cold pump regime. Parameters : \( T_\text{h} = 150\,\mathrm{K} \), \( T_\text{c} = 50\,\mathrm{K} \), and \( B_1 = 1\,\mathrm{T} \).}
    \label{fig9}
\end{figure}

\begin{figure}[!htbp]
    \centering
    \includegraphics[width=0.95\linewidth]{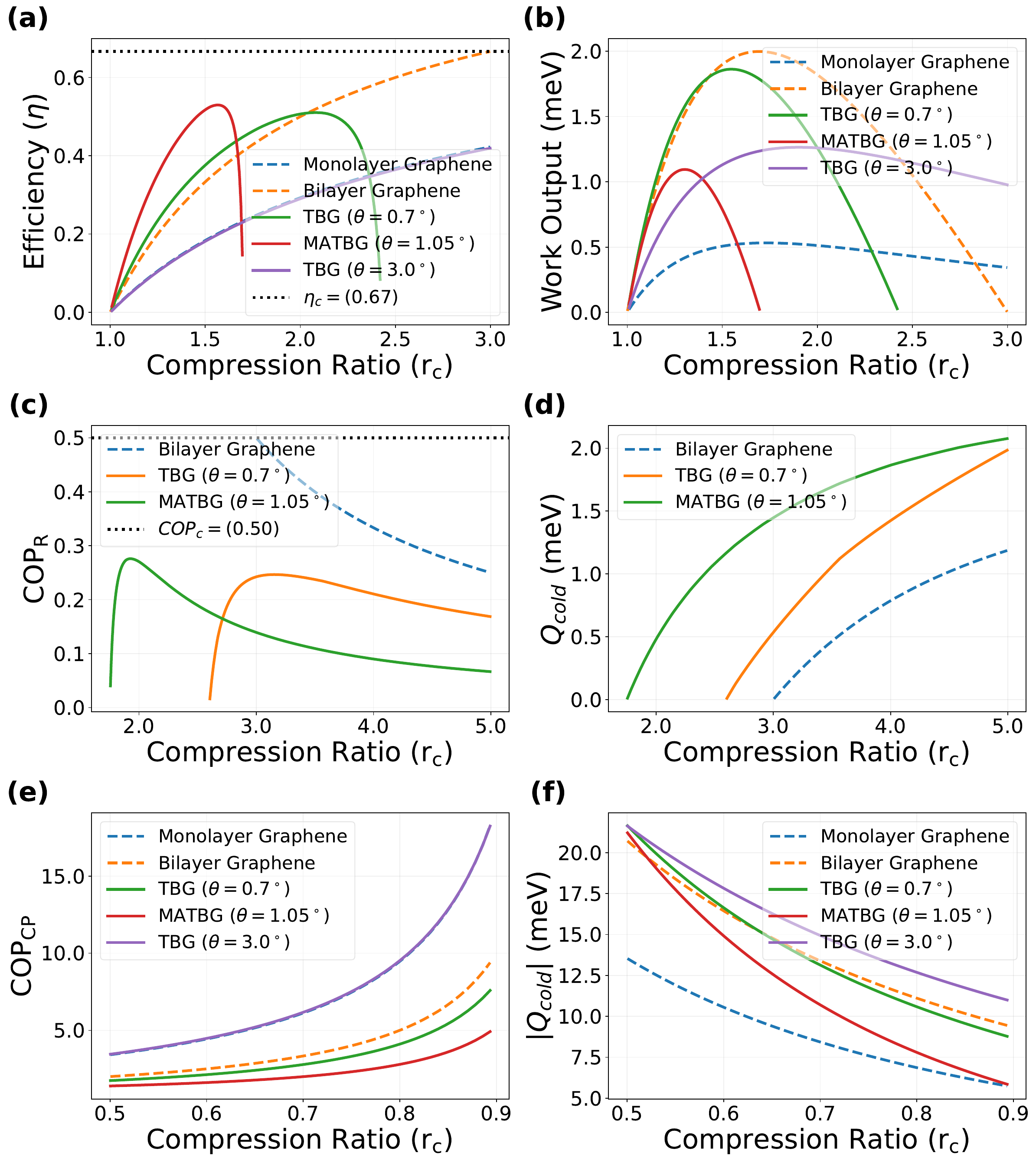}
    \caption{Performance of the QOC under strict adiabatic conditions as a function of the compression ratio \( r_c \) for different graphene systems. Subplots show (a) efficiency and (b) work output in the heat engine regime, (c) COP and (d) Refrigeration Output in the refrigerator regime, and (e) COP and (f) heat output in the cold pump regime. \( T_\text{h} = 150\,\mathrm{K} \), \( T_\text{c} = 50\,\mathrm{K} \), and \( B_1 = 1\,\mathrm{T} \).}
    \label{fig10}
\end{figure}

\subsection{Performance analysis with varying compression ratio}

Fig.~\ref{fig9} demonstrates the compression ratio ($r_c$) dependence of QOC performance across different graphene systems under the general adiabatic condition. In the heat engine phase (Fig.~\ref{fig9}(a),(b)), MATBG exhibits higher efficiency across the full range of compression ratios \(r_c\) compared to other graphene systems. However, the operational window in \(r_c\) for which MATBG functions as a heat engine is narrower. All systems approach Carnot efficiency as the compression ratio \(r_c\) increases; however, the work output vanishes in this limit. Fig~\ref{fig9}(c),(d) shows the performance of different graphene systems operating as refrigerators. All systems attain the Carnot COP (\(\text{COP}_\text{c}\)) at a lower value of \(r_c\) within the refrigeration regime, after which the COP decreases with increasing \(r_c\). MATBG exhibits a lower COP than other graphene systems at higher compression ratios \(r_c\); however, it remains within the refrigeration regime over a wider range of \(r_c\) values. The cold pump regime (Fig.~\ref{fig9}(e),(f)) shows MATBG underperforming again in COP despite its increasing trend with $r_c$.

\begin{figure}[!htbp]
    \centering
    \includegraphics[width=0.95\linewidth]{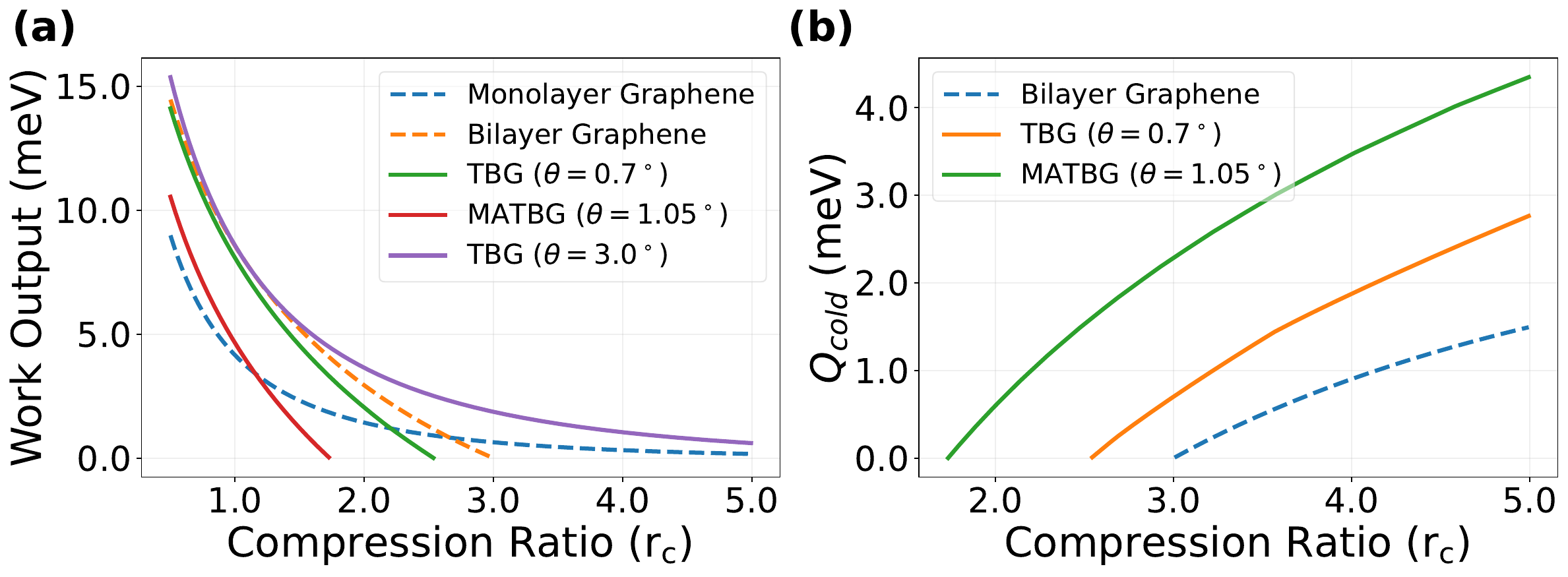}
    \caption{Performance of the QCC as a function of the compression ratio \( r_c \) for different graphene systems. Subplots show (a) work output in the heat engine regime where all systems operate with Carnot efficiency \(\eta_c=0.67\), (b) refrigeration output (\(Q_\text{cold}\)) in the refrigerator regime, where all systems operate with Carnot COP \(\text{COP}_c=0.5\). Parameters : \( T_\text{h} = 150\,\mathrm{K} \), \( T_\text{c} = 50\,\mathrm{K} \), and \( B_1 = 1\,\mathrm{T} \).}
    \label{fig11}

\end{figure}

\begin{figure}[!htbp]
    \centering
    \includegraphics[width=0.95\linewidth]{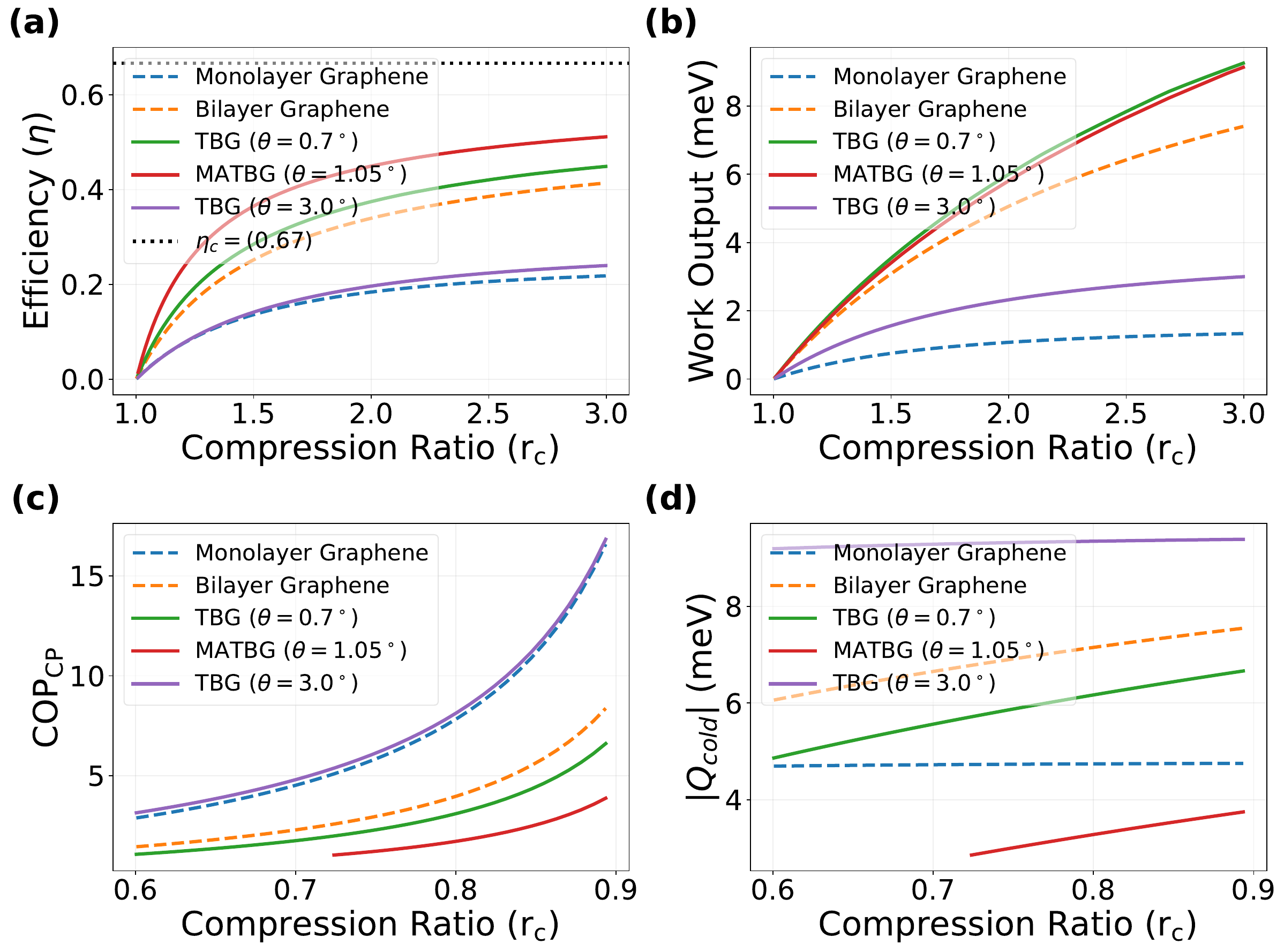}
    \caption{Performance of the QSC as a function of the compression ratio \( r_c \) for different graphene systems. Subplots show (a) efficiency and (b) work output in the heat engine regime, (c) COP and (d) heat output in the cold pump regime. Parameters : \( T_\text{h} = 150\,\mathrm{K} \), \( T_\text{c} = 50\,\mathrm{K} \), and \( B_1 = 1\,\mathrm{T} \).}
    \label{fig12}
\end{figure}

\begin{figure*}[!htbp]
    \centering
    \includegraphics[width=0.95\linewidth]{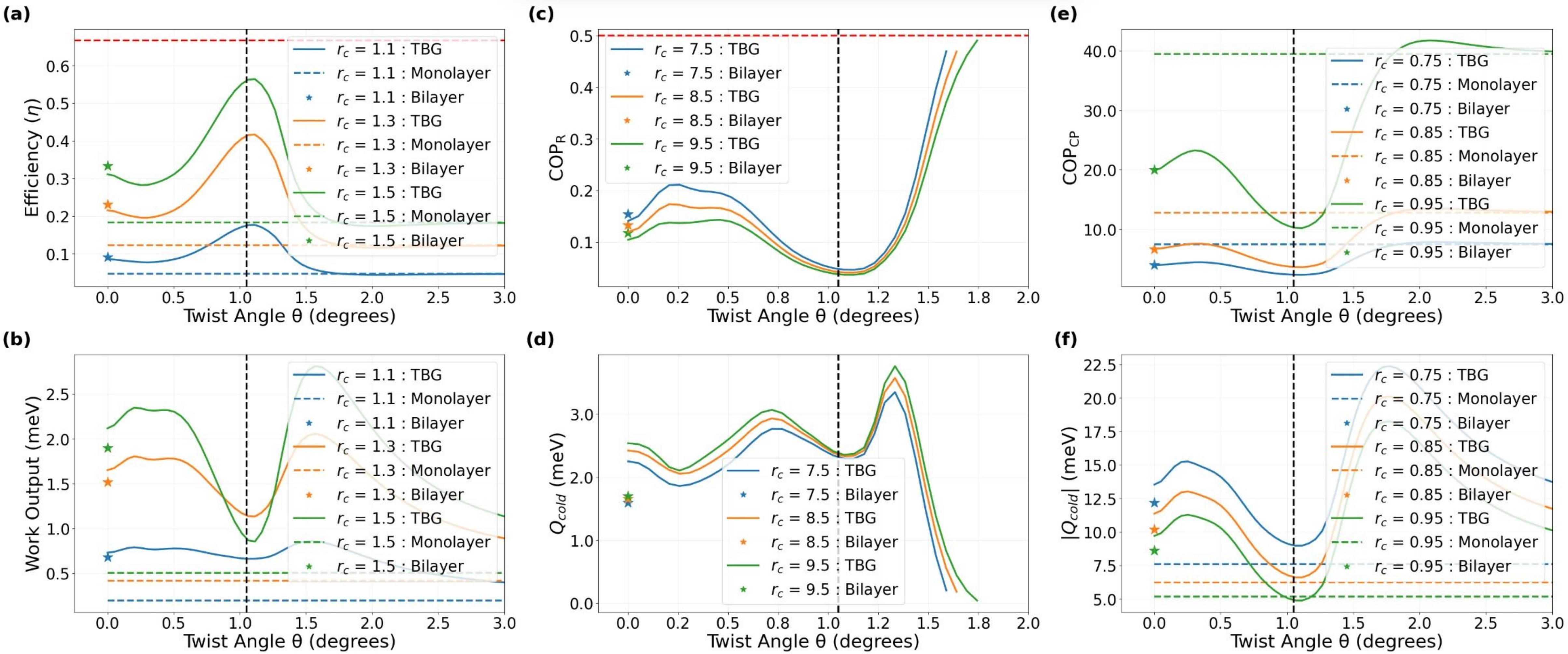}
    \caption{Performance of the QOC in general adiabatic conditions as a function of the Twist angle in the TBG system. Subplots show (a) efficiency with red dashed line indicating Carnot efficiency (\(\eta_c=0.67\)) and (b) work output in the heat engine regime, (c) COP with and red dashed line indicating Carnot COP (\(\text{COP}_c=0.50\))(d) refrigeration output in the refrigerator regime, and (e) COP and (f) heat output in the cold pump regime. In all subplots, the vertical black dashed line corresponds to the magic-angle (\(\theta=1.05^o\)). Parameters : \( T_\text{h} = 150\,\mathrm{K} \), \( T_\text{c} = 50\,\mathrm{K} \), and \( B_1 = 1\,\mathrm{T} \).}
    \label{fig13}
\end{figure*}

\begin{figure*}[!htbp]
    \centering
    \includegraphics[width=0.95\linewidth]{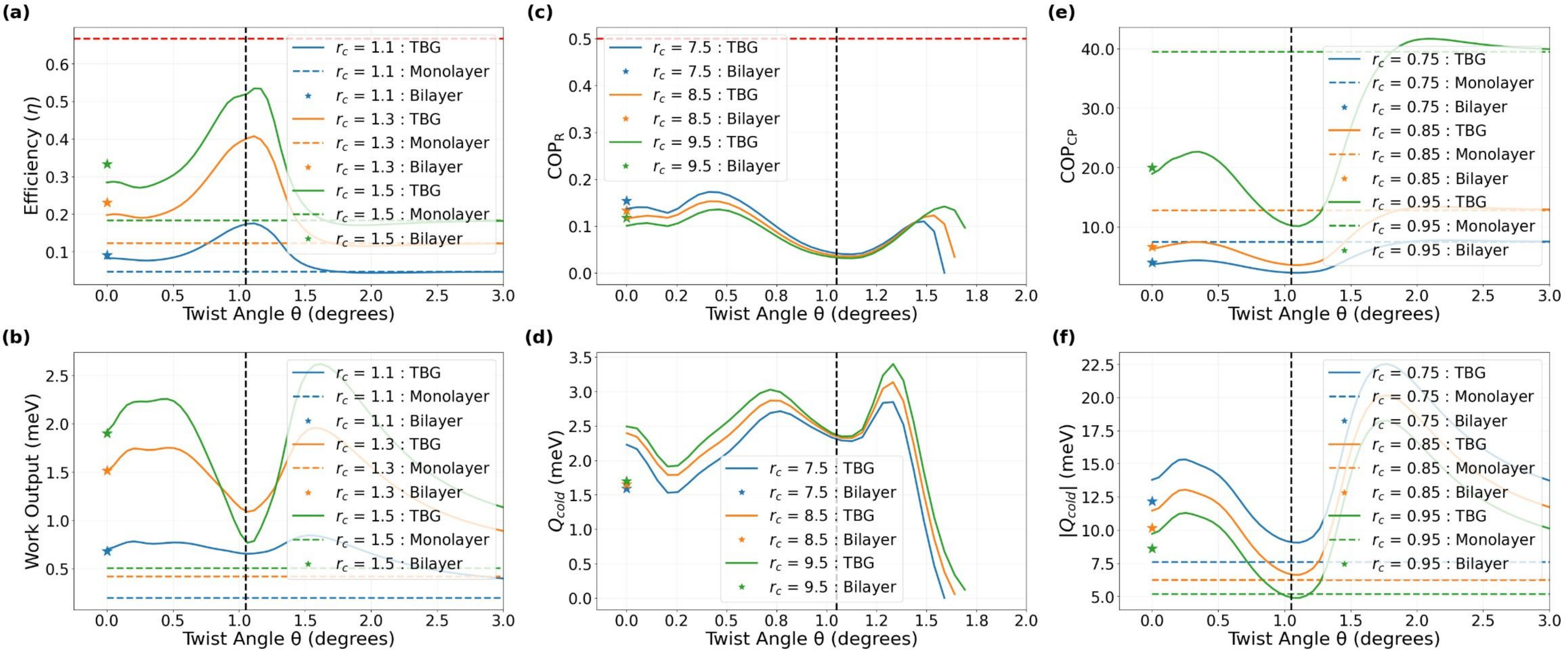}
    \caption{Performance of the QOC in strict adiabatic conditions as a function of the Twist angle in the TBG system. Subplots show (a) efficiency with red dashed line indicating Carnot efficiency (\(\eta_c=0.67\)) and (b) work output in the heat engine regime, (c) COP with and red dashed line indicating Carnot COP (\(\text{COP}_c=0.50\))(d) refrigeration output in the refrigerator regime, and (e) COP and (f) heat output in the cold pump regime. In all subplots, the vertical black dashed line corresponds to the magic-angle (\(\theta=1.05^o\)). Parameters : \( T_\text{h} = 150\,\mathrm{K} \), \( T_\text{c} = 50\,\mathrm{K} \), and \( B_1 = 1\,\mathrm{T} \).}
    \label{fig14}
\end{figure*}

Fig. \ref{fig10} illustrates the dependence of performance on the compression ratio \(r_c\) for the QOC under the strict adiabatic condition. Monolayer and bilayer graphene maintain identical performance to that of the general case. For TBG systems in heat engine phase, (Fig.~\ref{fig10}(a),(b)) efficiency initially increases with $r_c$ until reaching a sharp maximum, then decreasing rapidly. In the refrigeration phase (Fig.~\ref{fig10}(c)(d)), TBG systems exhibit an increase in COP to a maximum value, followed by a steady decrease as \(r_c\) is increased. The peak attained for efficiency in the heat engine phase and COP in the refrigeration phase is the largest for MATBG compared to other non-magic angle configurations and is attained at a lower \(r_c\) value. The cold pump regime (Fig.~\ref{fig10}(e),(f)) shows behavior similar to the general adiabatic case for all systems, with MATBG again displaying the lowest COP values. Strict adiabatic QOC also indicates the presence of a Joule pump phase where the system redistributes the work applied to it as heat to the external reservoirs.

One should note that as \( r_c \rightarrow 1 \), from Eqs.~(\ref{eq:otto heat1}),(\ref{eq:otto heat2}), we have \( Q_{\text{cold}}^{\text{gen}} = Q_{\text{cold}}^{\text{str}} \), \( Q_{\text{hot}}^{\text{gen}} = Q_{\text{hot}}^{\text{str}} \), and \( Q_{\text{cold}}^{\text{gen}} = -Q_{\text{hot}}^{\text{gen}} \). This implies that \( \text{W}^{\text{gen}} = \text{W}^{\text{str}} = 0 \), leading to \( \text{COP} \rightarrow \infty \) and \( \eta \rightarrow 0 \) \citep{lucio2025innovative}. Systems with larger twist angles ($\theta \approx 3^\circ$) converge toward monolayer graphene behavior.

\begin{figure}[!htbp]
    \centering
    \includegraphics[width=0.95\linewidth]{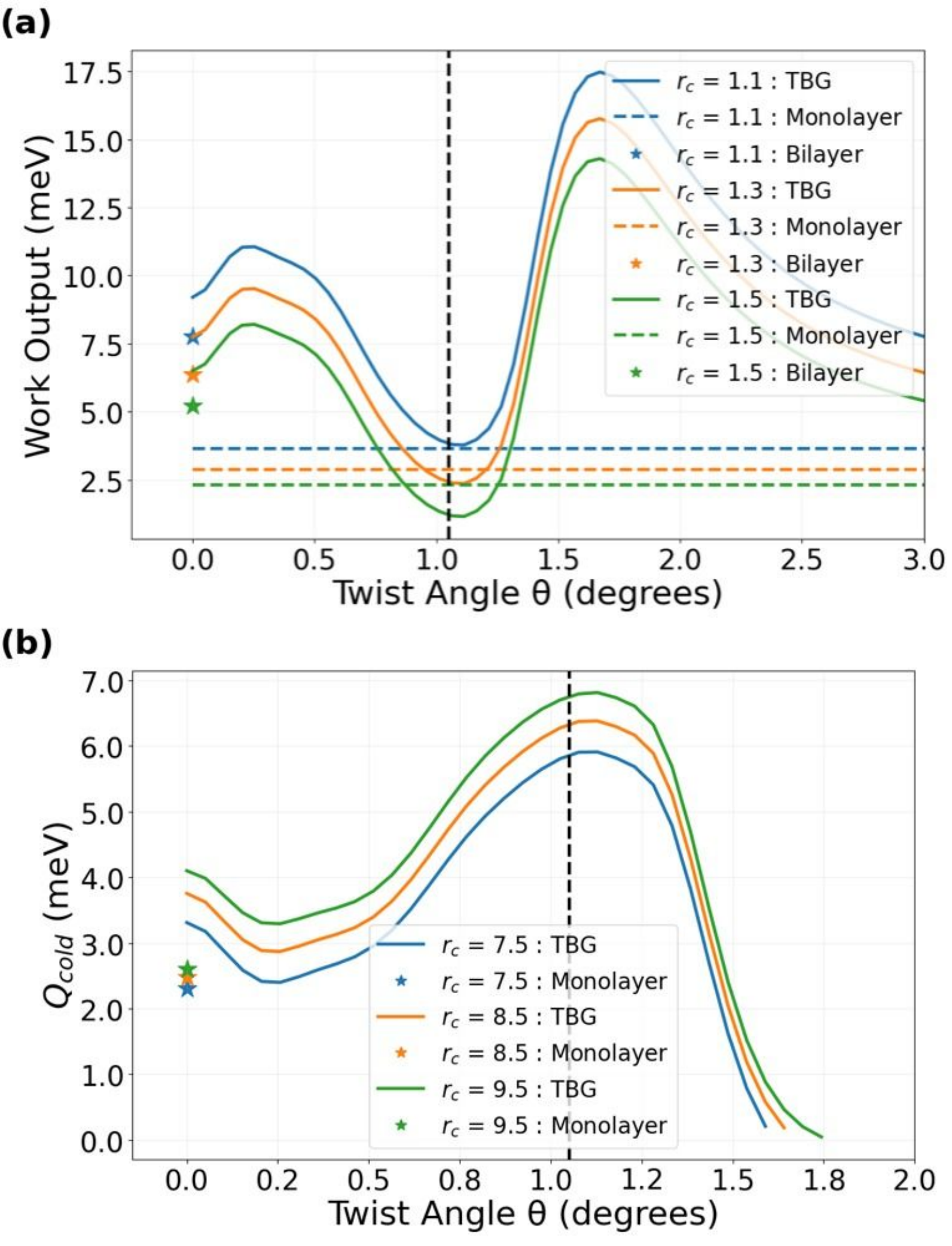}
    \caption{Performance of the QCC as a function of the Twist angle in the TBG system. Subplots show (a) work output in the heat engine regime, where all systems operate with Carnot efficiency (\(\eta_c =0.67\)) and (b) refrigeration output in the refrigeration regime, where all systems operate with Carnot COP (\(\text{COP}_c =0.50\)). In all subplots, the vertical black dashed line corresponds to the magic-angle (\(\theta=1.05^o\)). Parameters : \( T_\text{h} = 150\,\mathrm{K} \), \( T_\text{c} = 50\,\mathrm{K} \), and \( B_1 = 1\,\mathrm{T} \).}
    \label{fig15}
\end{figure}

Fig.~\ref{fig11} presents the performance of a QCC for different graphene systems as a function of $r_c$. All systems achieve the theoretical Carnot efficiency \(\eta_c\) limit in the heat engine regime due to reversibility, with work output varying between configurations. The compression ratio window, where MATBG is observed to be in the heat engine phase, is narrow, and it underperforms compared to the other systems. The refrigeration regime also shows all systems operating at the Carnot COP (\(\text{COP}_c\)). The refrigeration output \(Q_{\text{cold}}\) is highest for MATBG compared to other systems.

Fig~\ref{fig12} shows the performance of the QSC as a function of \(r_c\). In the heat engine phase (Fig.~\ref{fig12}(a),(b)), MATBG exhibits the highest efficiency and a work output comparable to or greater than other systems. Furthermore, as the compression ratio increases, the system remains in the heat engine regime, with both efficiency and work output continuing to increase. In the cold pump phase (Fig.~\ref{fig12}(c),(d)), MATBG underperforms with the lowest COP and heat dumped ($|Q_\text{cold}|$) compared to other systems. Just like QOC, here as \(r_c \rightarrow 1\), using Eqs.(\ref{eq:sterlingheat1}), (\ref{eq:sterlingheat2}),(\ref{eq:sterlingheat3}),(\ref{eq:sterlingheat4}), we find that $Q_{AB}=Q_{CD}=0$ and $Q_{BC}=-Q_{DA}$, which implies $Q_\text{hot}=-Q_\text{cold}$, hence $\text{W}=0$, which gives us $COP \rightarrow \infty$ and $\eta \rightarrow 0$ \citep{lucio2025innovative}.

\begin{figure}[!htbp]
    \centering
    \includegraphics[width=0.93\linewidth]{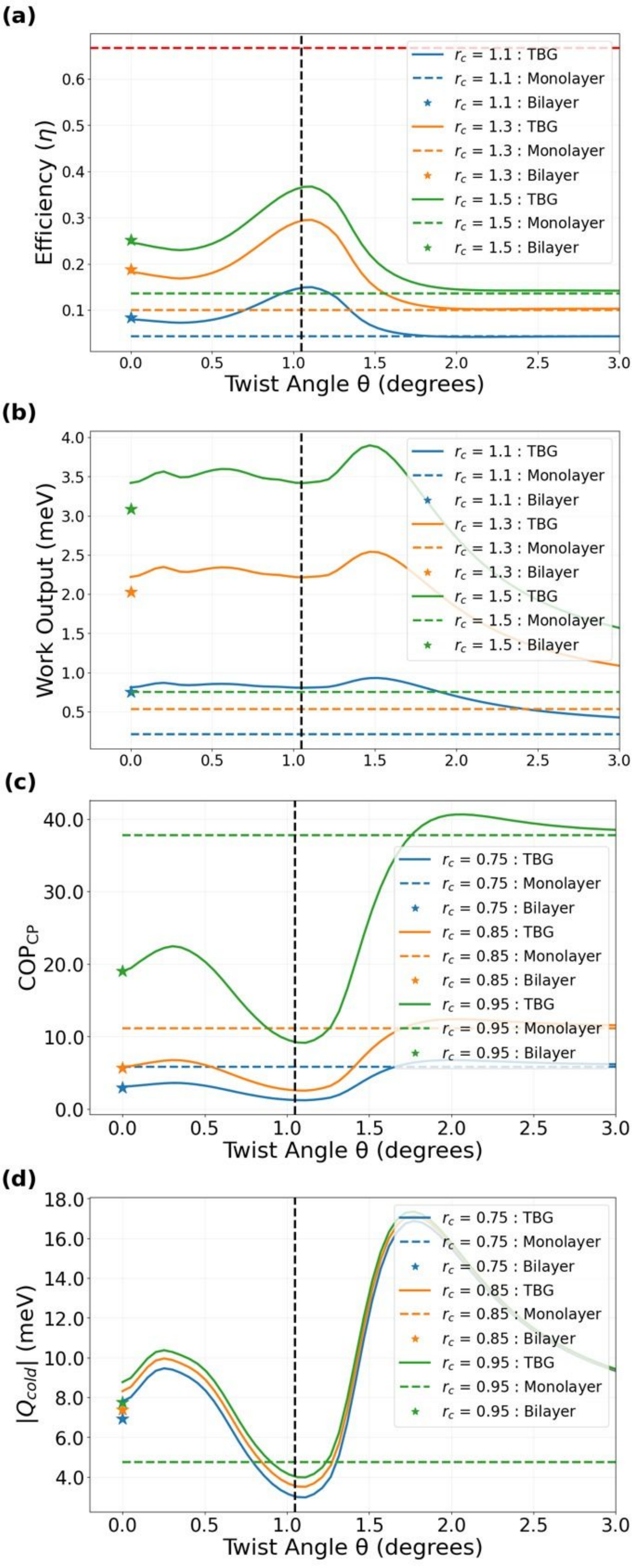}
    \caption{Performance of the QSC as a function of the Twist angle in the TBG system. Subplots show (a) efficiency with horizontal red dashed line indicating Carnot efficiency (\( \eta_c=0.67 \) ) and (b) work output in the heat engine regime, (c) COP, and (d) heat output in the cold pump regime. The heat output for monolayer graphene remains constant irrespective of the compression ratio value. In all subplots, the vertical black dashed line corresponds to the magic-angle (\(\theta=1.05^o\)). Parameters : \( T_\text{h} = 150\,\mathrm{K} \), \( T_\text{c} = 50\,\mathrm{K} \), and \( B_1 = 1\,\mathrm{T} \).}
    \label{fig16}
\end{figure}

\subsection{Performance analysis with varying twist angles}


Figs.~\ref{fig13} and~\ref{fig14} illustrate the performance of the QOC under general and strict adiabatic conditions respectively, as a function of the twist angle for selected compression ratio values. The heat engine phase (Fig.~\ref{fig13}(a) and Fig.~\ref{fig14}(a)) shows a peak in efficiency near the magic angle. However, this peak coincides with a notable reduction in work output as seen in Fig.~\ref{fig13}(b) and Fig.~\ref{fig14}(b) for higher \(r_c\). In the refrigeration and cold pump regimes (Fig.~\ref{fig13}(c),(e) and Fig.~\ref{fig14}(c),(e)), the COP is significantly reduced near the magic angle, indicating MATBG’s poor performance for these modes compared to other twist angles. This trend holds under both general and strict adiabatic conditions.


For the QCC (Fig.~\ref{fig15}(a)), the heat engine regime has a reduced work output near the magic angle, then increases to a peak before gradually declining with larger twist angles. In the refrigeration phase (Fig.~\ref{fig15}(b)), although all systems operate at Carnot COP, MATBG has the highest refrigeration output (\(Q_\text{cold}\)) compared to other twist angles.


\begin{figure}[!htbp]
    \centering
    \includegraphics[width=0.95\linewidth]{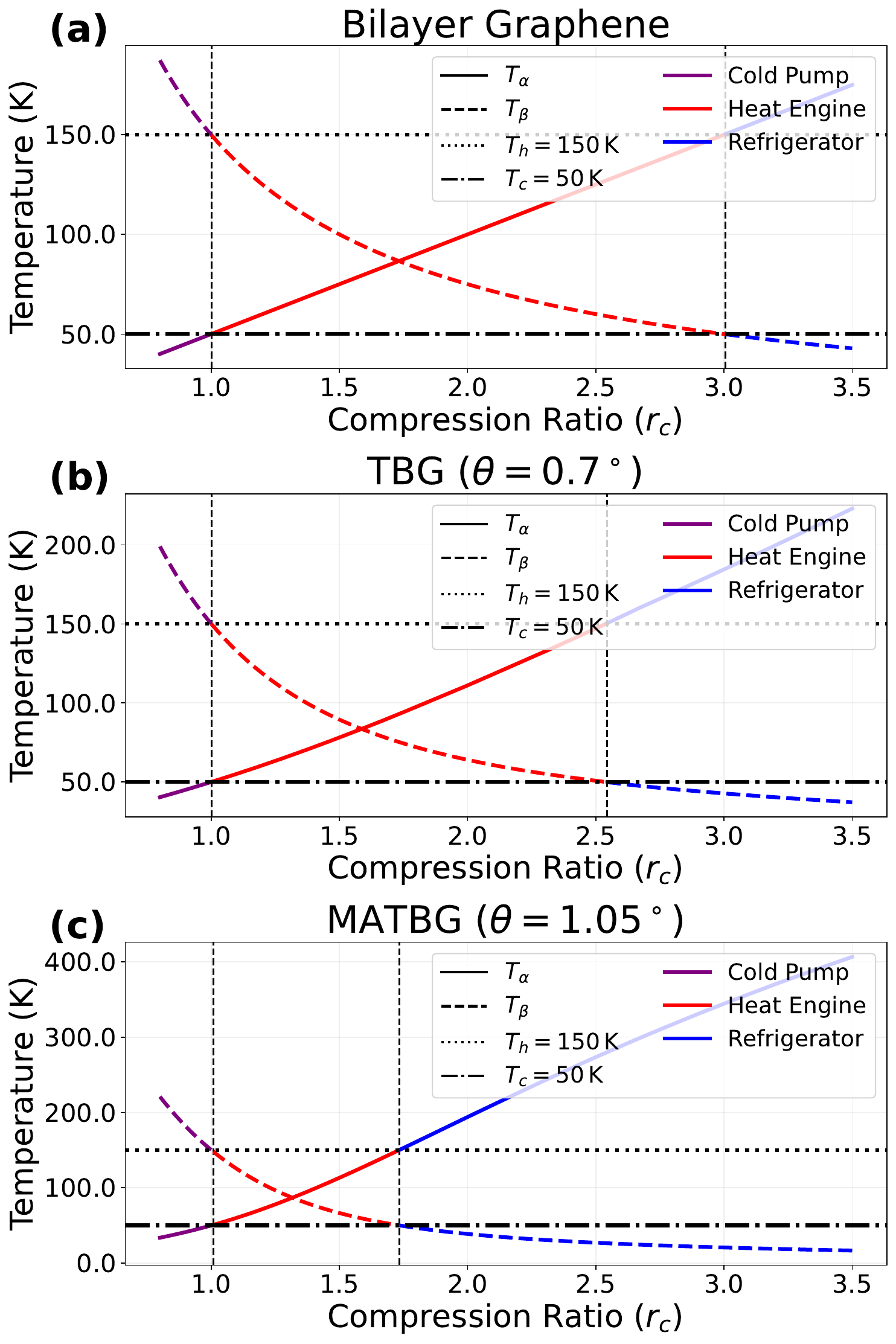}
    \caption{{Effective temperature (\(T_\alpha\) and \(T_\beta\))  as a function of compression ratio at the end of the adiabatic stroke for a QOC under general adiabatic conditions. The plots show results for (a) bilayer graphene, (b) TBG at \(\theta = \ang{0.7}\), and (c)MATBG (\(\theta = \ang{1.05}\)). The vertical dashed line depicts transition. Parameters: \(T_h = \SI{150}{K}\), \(T_c = \SI{50}{K}\), and \(B_1 = \SI{1}{T}\).}}
    \label{fig17}
\end{figure}


Fig.~\ref{fig16} presents the performance of TBG as a function of the twist angle for graphene systems in QSC. In the heat engine phase, the efficiency follows a trend similar to the QOC, though slightly lower at the same \(r_c\)(Fig.~\ref{fig16}(a)); however the work output is not suppressed near the magic angle as seen in QOC. The QSC yields nearly constant work output at small angles, with both efficiency and work decreasing at larger angles (Fig.~\ref{fig16}(a),(b)). In the cold pump phase (Fig.~\ref{fig16}(c),(d)), the COP trend resembles that of the QOC but with lower heat output to the cold reservoir (\(Q_\text{cold}\)) near the magic angle. The heat output into the cold reservoir (\(|Q_\text{cold}|\)) is found to be constant for monolayer and TBG at larger twist angles.

\begin{figure}[!htbp]
    \centering
    \includegraphics[width=0.95\linewidth]{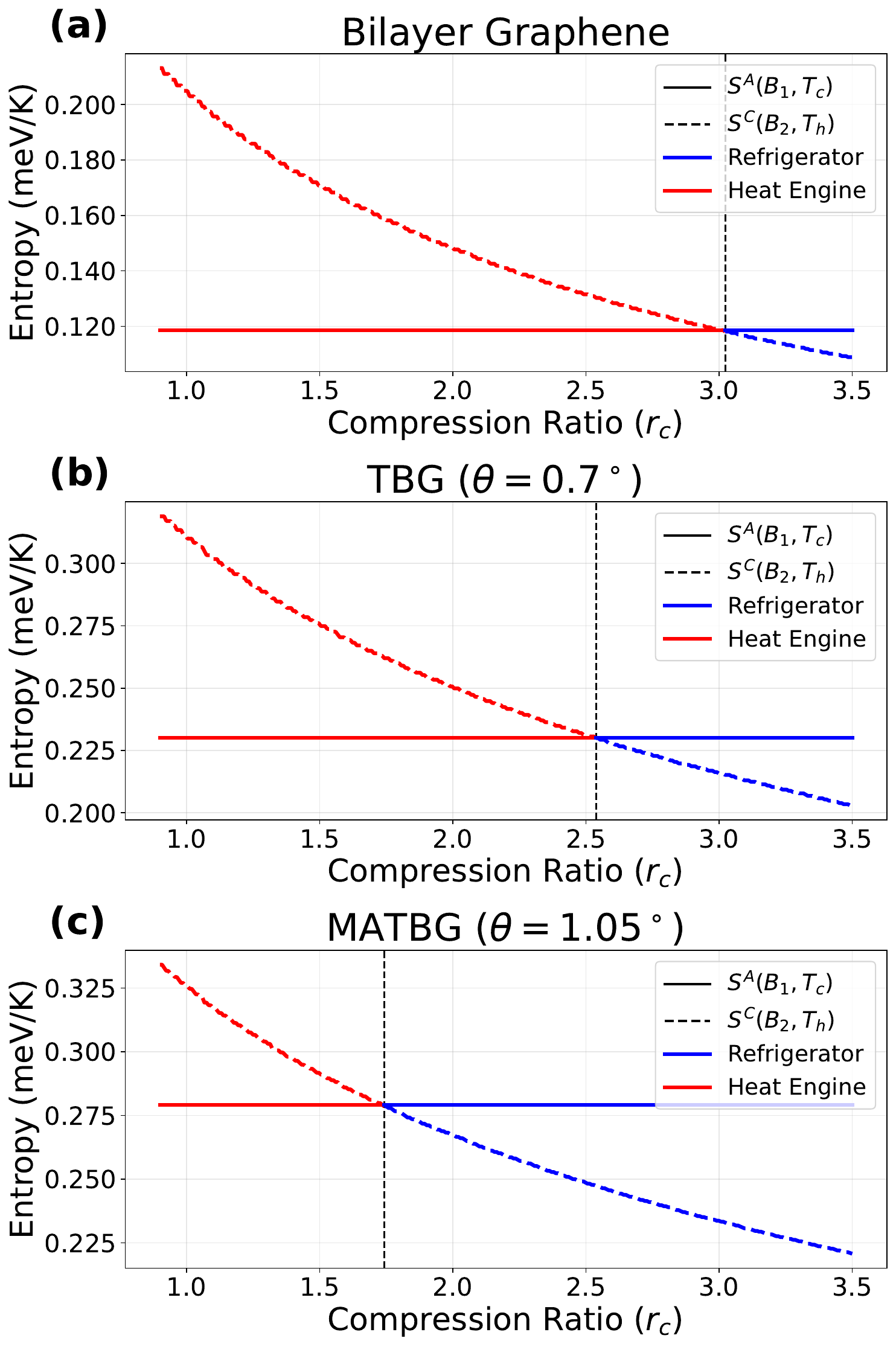}
    \caption{{Entropy at points A and C (\(S^A\) and \(S^C\)) for a QCC as a function of compression ratio. The plots show results for (a) bilayer graphene, (b) TBG at \(\theta = \ang{0.7}\), and (c)MATBG (\(\theta = \ang{1.05}\)). The vertical dashed line depicts transition. Parameters: \(T_h = \SI{150}{K}\), \(T_c = \SI{50}{K}\), and \(B_1 = \SI{1}{T}\).}}
    \label{fig18}
\end{figure}

\section{Analysis}
\label{sec:analys}


In this section, we analyze the results obtained in the previous section and compare the performance of different operational phases across different cycles.

{

\subsection{Operational phase regimes and transition condition}

The operational phase transitions in graphene based systems undergoing a general QOC are governed by the adiabatic conditions in Eqs.~\eqref{eq:ottoaida1} and \eqref{eq:ottoaida2}. Depending on the effective temperatures attained at the end of the adiabatic strokes, \(T_\alpha\) and \(T_\beta\), the system can exist in distinct operational phases. Fig.~\ref{fig17} shows the variation of these effective temperatures with the compression ratio for different graphene systems. For compression ratios \(r_c \le 1\), all systems satisfy \(T_\alpha \le T_c \le T_h \le T_\beta\), corresponding to the cold-pump phase. As the compression ratio increases, the system transitions into a heat-engine phase, characterized by \(T_c \le T_\alpha, T_\beta \le T_h\). A second transition to refrigeration phase occurs when the entropies satisfy \(S^C(B_2,T_h) = S^A(B_1,T_c)\), with further increase in compression ratio the effective temperatures obey \(T_\beta \le T_c \le T_h \le T_\alpha\). Although the overall phase structure depends on the specific cycle configuration, the compression ratio range over which a given graphene system operates and its overall performance in each phase is determined by how its entropy varies with magnetic field. Since MATBG undergoes the second transition at a smaller compression ratio than other graphene systems, its entropy exhibits a stronger dependence on the magnetic field. 

For a strict QOC, the transition condition for the bilayer graphene remains identical to Fig.~\ref{fig17}(a), as the energy-scaling condition (see, Appendix C) ensures the system stays in a thermal state with a well-defined effective temperature. However, for TBG, the strict adiabatic process drives the system out of thermal equilibrium, leading to a non-thermal state with no well-defined temperature\citep{pena2020otto}. This gives rise to additional operational regimes, such as the Joule pump phase, a unique feature of TBG arising from its energy levels not scaling uniformly with the magnetic field as seen in Fig.\ref{fig6}(b).}

\begin{figure}[!htbp]
    \centering
    \includegraphics[width=0.95\linewidth]{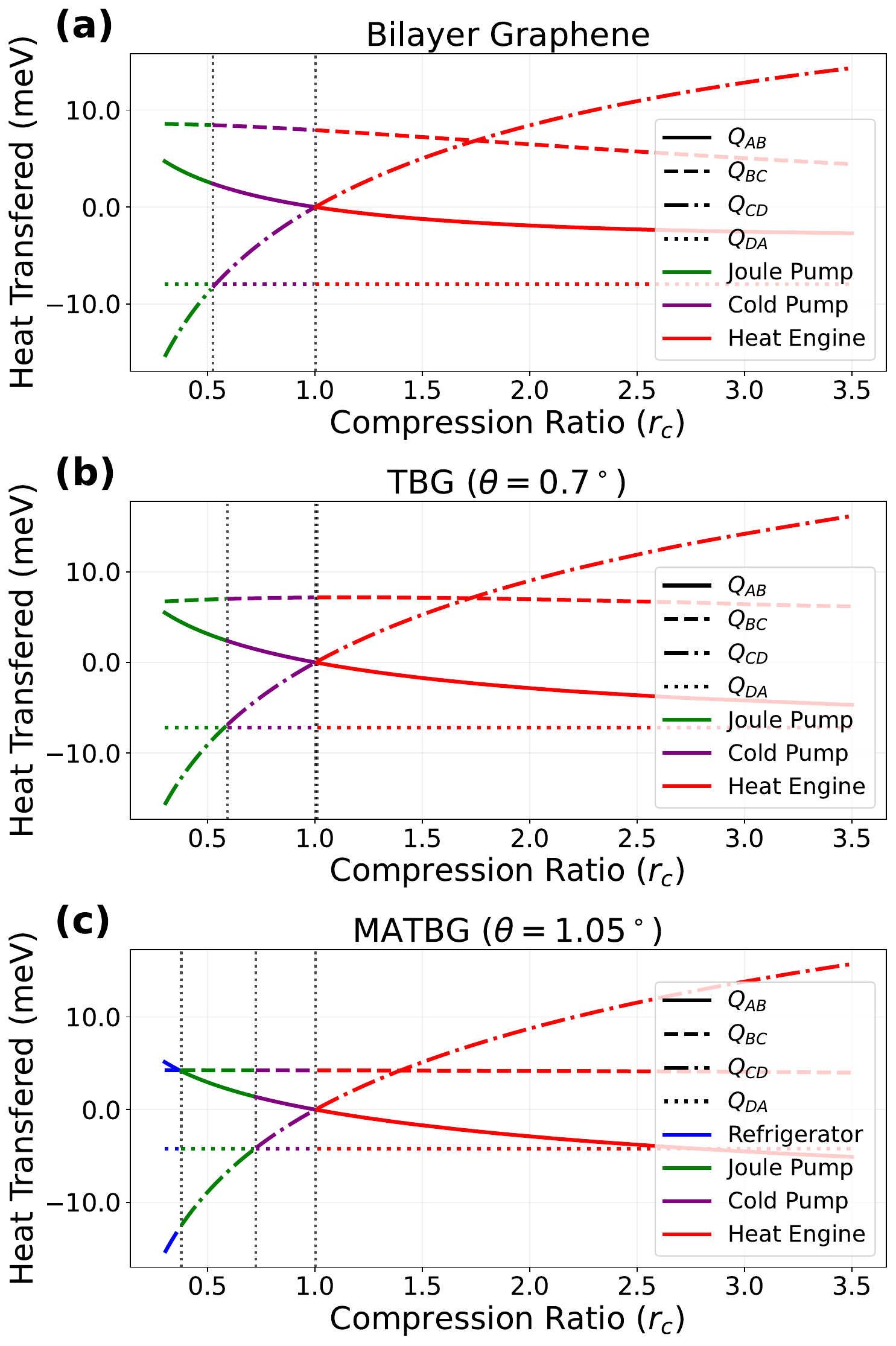}
    \caption{{Heat transferred during individual strokes for a QSC as a function of compression ratio. The plots show results for (a) bilayer graphene, (b) TBG at \(\theta = \ang{0.7}\), and (c)MATBG (\(\theta = \ang{1.05}\)). The vertical dashed line depicts transition. Parameters: \(T_h = \SI{150}{K}\), \(T_c = \SI{50}{K}\), and \(B_1 = \SI{1}{T}\).}}
    \label{fig19}
\end{figure}

{
For the QCC, the operational phase transition is determined by the condition 
\(S^A(B_1,T_c) = S^C(B_2,T_h)\) as seen in Fig.\ref{fig18}. Systems operate in the heat engine phase when \(S^A(B_1,T_c) < S^C(B_2,T_h)\), and in the refrigerator phase when  \(S^A(B_1,T_c) > S^C(B_2,T_h)\). Since \(B_1\), \(T_h\), and \(T_c\) are fixed parameters,  \(S^A(B_1,T_c)\) is independent of the compression ratio, while \(S^C(B_2,T_h)\) decreases as the compression ratio increases. The entropy of MATBG at a given compression ratio and temperature is higher than that of other graphene systems, which can be attributed to its flat band dispersion, as it leads to closely spaced Landau levels and high occupation probabilities for higher Landau levels (see, Appendix B). This causes thermal effects to be less pronounced, causing the operational phase transition for MATBG to occur at a lower compression ratio compared to other graphene based systems.
}

\begin{table*}[!htbp]
\begin{tabular}{|c|c|ccc|ccc|c|ccc|}
\hline
\multirow{2}{*}{\textbf{\begin{tabular}[c]{@{}c@{}}Compression\\ Ratio Value\end{tabular}}} & \multirow{2}{*}{\textbf{\begin{tabular}[c]{@{}c@{}}Graphene\\ System\end{tabular}}} & \multicolumn{3}{c|}{\textbf{\begin{tabular}[c]{@{}c@{}}General QOC\end{tabular}}}                                                                                                                                                      & \multicolumn{3}{c|}{\textbf{\begin{tabular}[c]{@{}c@{}}Strict QOC\end{tabular}}}                                                                                                                                              & \textbf{QCC}                                & \multicolumn{3}{c|}{\textbf{QSC}}                                                                                                                                                                                                         \\ \cline{3-12} 
                                                                                            &                                                                                     & \multicolumn{1}{c|}{\textbf{\begin{tabular}[c]{@{}c@{}} Output W \\ (meV)\end{tabular}}} & \multicolumn{1}{c|}{\textbf{\begin{tabular}[c]{@{}c@{}} \(\frac{\eta}{\eta_c}\)\end{tabular}}} & \textbf{\boldmath\(\text{W} \times \frac{\eta}{\eta_c}\)}& \multicolumn{1}{c|}{\textbf{\begin{tabular}[c]{@{}c@{}} Output W \\ (meV)\end{tabular}}} & \multicolumn{1}{c|}{\textbf{\begin{tabular}[c]{@{}c@{}} \(\frac{\eta}{\eta_c}\)\end{tabular}}} &\textbf{\boldmath\(\text{W} \times \frac{\eta}{\eta_c}\)}& \textbf{\boldmath\(\text{W} \times \frac{\eta}{\eta_c}\)}& \multicolumn{1}{c|}{\textbf{\begin{tabular}[c]{@{}c@{}} Output W \\ (meV)\end{tabular}}} & \multicolumn{1}{c|}{\textbf{\begin{tabular}[c]{@{}c@{}} \(\frac{\eta}{\eta_c}\)\end{tabular}}} & \textbf{\boldmath\(\text{W} \times \frac{\eta}{\eta_c}\)}\\ \hline
\multirow{5}{*}{\textbf{\(\text{r}_{\text{c}}\) = 1.1}}                                                          & \textbf{Monolayer}                                                                  & \multicolumn{1}{c|}{0.199}                                                                    & \multicolumn{1}{c|}{0.071}                                                                  & 0.014                                       & \multicolumn{1}{c|}{0.199}                                                                    & \multicolumn{1}{c|}{0.071}                                                                  & 0.014                              & 3.885                                       & \multicolumn{1}{c|}{0.217}                                                                    & \multicolumn{1}{c|}{0.065}                                                                  & 0.014                                       \\ \cline{2-12} 
                                                                                            & \textbf{Bilayer}                                                                    & \multicolumn{1}{c|}{0.689}                                                                    & \multicolumn{1}{c|}{0.138}                                                                  & 0.095                                       & \multicolumn{1}{c|}{0.689}                                                                    & \multicolumn{1}{c|}{0.138}                                                                  & 0.095                              & 7.749                                       & \multicolumn{1}{c|}{0.762}                                                                    & \multicolumn{1}{c|}{0.217}                                                                  & 0.096                                       \\ \cline{2-12} 
                                                                                            & \textbf{TBG (\(0.7^o\))}                                                                  & \multicolumn{1}{c|}{\color[HTML]{3197FF} \textbf{0.746}}                                                                    & \multicolumn{1}{c|}{0.166}                                                                  & 0.124                                       & \multicolumn{1}{c|}{\color[HTML]{3197FF} \textbf{0.741}}                                                                    & \multicolumn{1}{c|}{0.165}                                                                  & 0.122                              & 7.233                                       & \multicolumn{1}{c|}{\color[HTML]{3197FF} \textbf{0.841}}                                                                    & \multicolumn{1}{c|}{0.150}                                                                  & 0.126                                       \\ \cline{2-12} 
                                                                                            & \textbf{MATBG (\(1.05^o\))}                                                               & \multicolumn{1}{c|}{0.663}                                                                    & \multicolumn{1}{c|}{\color[HTML]{3197FF} \textbf{0.263}}                                                                  & \color[HTML]{3197FF} \textbf{0.174}                              & \multicolumn{1}{c|}{0.655}                                                                    & \multicolumn{1}{c|}{\color[HTML]{3197FF} \textbf{0.260}}                                                                  & \color[HTML]{3197FF} \textbf{0.170}                     & 3.824                                       & \multicolumn{1}{c|}{0.806}                                                                    & \multicolumn{1}{c|}{\color[HTML]{3197FF} \textbf{0.222 }}                                                                  & \color[HTML]{3197FF} \textbf{0.179}                              \\ \cline{2-12} 
                                                                                            & \textbf{TBG (\(3.0^o\))}                                                                  & \multicolumn{1}{c|}{0.398}                                                                    & \multicolumn{1}{c|}{0.069}                                                                  & 0.027                                       & \multicolumn{1}{c|}{0.398}                                                                    & \multicolumn{1}{c|}{0.069}                                                                  & 0.027                              & \color[HTML]{3197FF} \textbf{7.752}                              & \multicolumn{1}{c|}{0.426}                                                                    & \multicolumn{1}{c|}{0.065}                                                                  & 0.028                                       \\ \hline
\multirow{5}{*}{\textbf{\(\text{r}_{\text{c}}\) = 1.35}}                                                         & \textbf{Monolayer}                                                                  & \multicolumn{1}{c|}{0.447}                                                                    & \multicolumn{1}{c|}{0.208}                                                                  & 0.093                                       & \multicolumn{1}{c|}{0.447}                                                                    & \multicolumn{1}{c|}{0.208}                                                                  & 0.093                              & 2.947                                       & \multicolumn{1}{c|}{0.593}                                                                    & \multicolumn{1}{c|}{0.166}                                                                  & 0.098                                       \\ \cline{2-12} 
                                                                                            & \textbf{Bilayer}                                                                    & \multicolumn{1}{c|}{1.642}                                                                    & \multicolumn{1}{c|}{0.387}                                                                  & 0.636                                       & \multicolumn{1}{c|}{1.642}                                                                    & \multicolumn{1}{c|}{0.387}                                                                  & 0.636                              & 6.113                                       & \multicolumn{1}{c|}{2.304}                                                                    & \multicolumn{1}{c|}{0.309}                                                                  & 0.712                                       \\ \cline{2-12} 
                                                                                            & \textbf{TBG (\(0.7^o\))}                                                                  & \multicolumn{1}{c|}{\color[HTML]{3197FF} \textbf{1.727}}                                                                    & \multicolumn{1}{c|}{0.457}                                                                  & \color[HTML]{3197FF} \textbf{0.789}                              & \multicolumn{1}{c|}{\color[HTML]{3197FF} \textbf{1.681}}                                                                    & \multicolumn{1}{c|}{0.448}                                                                  & \color[HTML]{3197FF} \textbf{0.753}                     & 5.470                                       & \multicolumn{1}{c|}{\color[HTML]{3197FF} \textbf{2.613}}                                                                    & \multicolumn{1}{c|}{0.354}                                                                  & 0.926                                       \\ \cline{2-12} 
                                                                                            & \textbf{MATBG (\(1.05^o\))}                                                               & \multicolumn{1}{c|}{1.138}                                                                    & \multicolumn{1}{c|}{\color[HTML]{3197FF} \textbf{0.681}}                                                                  & 0.775                                       & \multicolumn{1}{c|}{1.077}                                                                    & \multicolumn{1}{c|}{\color[HTML]{3197FF} \textbf{0.654}}                                                                  & 0.704                              & 2.115                                       & \multicolumn{1}{c|}{2.51}                                                                     & \multicolumn{1}{c|}{\color[HTML]{3197FF} \textbf{0.471 }}                                                                  & \color[HTML]{3197FF} \textbf{1.183}                              \\ \cline{2-12} 
                                                                                            & \textbf{TBG (\(3.0^o\))}                                                                  & \multicolumn{1}{c|}{0.962}                                                                    & \multicolumn{1}{c|}{0.205}                                                                  & 0.198                                       & \multicolumn{1}{c|}{0.962}                                                                    & \multicolumn{1}{c|}{0.205}                                                                  & 0.197                              & \color[HTML]{3197FF} \textbf{6.163}                              & \multicolumn{1}{c|}{1.211}                                                                    & \multicolumn{1}{c|}{0.170}                                                                  & 0.206                                       \\ \hline
\multirow{5}{*}{\textbf{\(\text{r}_{\text{c}}\) = 1.6}}                                                          & \textbf{Monolayer}                                                                  & \multicolumn{1}{c|}{0.526}                                                                    & \multicolumn{1}{c|}{0.315}                                                                  & 0.166                                       & \multicolumn{1}{c|}{0.526}                                                                    & \multicolumn{1}{c|}{0.315}                                                                  & 0.166                              & 2.260                                       & \multicolumn{1}{c|}{0.84}                                                                     & \multicolumn{1}{c|}{0.225}                                                                  & 0.189                                       \\ \cline{2-12} 
                                                                                            & \textbf{Bilayer}                                                                    & \multicolumn{1}{c|}{\color[HTML]{3197FF} \textbf{\color[HTML]{3197FF} \textbf{1.978}}}                                                                    & \multicolumn{1}{c|}{0.564}                                                                  & 1.115                                       & \multicolumn{1}{c|}{\color[HTML]{3197FF} \textbf{1.978}}                                                                    & \multicolumn{1}{c|}{0.564}                                                                  & 1.115                              & 4.698                                       & \multicolumn{1}{c|}{3.559}                                                                    & \multicolumn{1}{c|}{0.414}                                                                  & 1.474                                       \\ \cline{2-12} 
                                                                                            & \textbf{TBG (\(0.7^o\))}                                                                  & \multicolumn{1}{c|}{1.963}                                                                    & \multicolumn{1}{c|}{0.652}                                                                  & \color[HTML]{3197FF} \textbf{1.28}                               & \multicolumn{1}{c|}{1.851}                                                                    & \multicolumn{1}{c|}{0.625}                                                                  & \color[HTML]{3197FF} \textbf{1.157}                     & 3.966                                       & \multicolumn{1}{c|}{\color[HTML]{3197FF} \textbf{4.112}}                                                                    & \multicolumn{1}{c|}{0.465 }                                                                  & 1.911                                       \\ \cline{2-12} 
                                                                                            & \textbf{MATBG (\(1.05^o\))}                                                               & \multicolumn{1}{c|}{0.562}                                                                    & \multicolumn{1}{c|}{\color[HTML]{3197FF} \textbf{0.919}}                                                                  & 0.516                                       & \multicolumn{1}{c|}{0.432}                                                                    & \multicolumn{1}{c|}{\color[HTML]{3197FF} \textbf{0.785}}                                                                  & 0.34                               & 0.658                                       & \multicolumn{1}{c|}{3.962}                                                                    & \multicolumn{1}{c|}{\color[HTML]{3197FF} \textbf{0.585}}                                                                  & \color[HTML]{3197FF} \textbf{2.317}                              \\ \cline{2-12} 
                                                                                            & \textbf{TBG (\(3.0^o\))}                                                                  & \multicolumn{1}{c|}{1.198}                                                                    & \multicolumn{1}{c|}{0.312}                                                                  & 0.373                                       & \multicolumn{1}{c|}{1.198}                                                                    & \multicolumn{1}{c|}{0.312}                                                                  & 0.373                              & \color[HTML]{3197FF} \textbf{4.961}                              & \multicolumn{1}{c|}{1.762}                                                                    & \multicolumn{1}{c|}{0.235}                                                                  & 0.414                                       \\ \hline
\end{tabular}
\caption{Efficiency, Work output in (meV) and coefficient of merit (\(W \times \frac{\eta}{\eta_c}\)) for QOC (general and strict), QCC, QSC for specific compression ratio values operating in \(\textbf{Heat Engine}\) phase. The efficiencies are in units of Carnot efficiency (\(\eta_c=0.67\)).  \( T_\text{h} = 150\,\mathrm{K} \), \( T_\text{c} = 50\,\mathrm{K} \), and \( B_1 = 1\,\mathrm{T} \).}
\label{tab:he_comparison}
\end{table*}

\begin{table*}[!htbp]
\begin{tabular}{|c|c|ccc|ccc|c|ccc|}
\hline
\multirow{2}{*}{\textbf{\begin{tabular}[c]{@{}c@{}}Compression\\ Ratio Value\end{tabular}}} & \multirow{2}{*}{\textbf{\begin{tabular}[c]{@{}c@{}}Graphene\\ System\end{tabular}}} & \multicolumn{3}{c|}{\textbf{\begin{tabular}[c]{@{}c@{}}General QOC\end{tabular}}}                                                                                                                                              & \multicolumn{3}{c|}{\textbf{\begin{tabular}[c]{@{}c@{}}Strict QOC\end{tabular}}}                                                                                                                                               & \textbf{QCC}    &  \multicolumn{3}{c|}{\textbf{\begin{tabular}[c]{@{}c@{}}QSC\end{tabular}}}   \\ \cline{3-12} 
                                                                                            &                                                                                     & \multicolumn{1}{c|}{\textbf{\begin{tabular}[c]{@{}c@{}} Output \\ \(Q_\text{cold}\)\\ (meV)\end{tabular}}} & \multicolumn{1}{c|}{\textbf{\begin{tabular}[c]{@{}c@{}} \(\frac{\text{COP}_\text{R}}{\text{COP}_\text{c}}\)\end{tabular}}} & \(Q_{\text{cold}} \times \frac{\text{COP}_\text{R}}{\text{COP}_\text{c}}\) & \multicolumn{1}{c|}{\textbf{\begin{tabular}[c]{@{}c@{}} Output \\  \(Q_\text{cold}\)\\ (meV)\end{tabular}}} & \multicolumn{1}{c|}{\textbf{\begin{tabular}[c]{@{}c@{}} \(\frac{\text{COP}_\text{R}}{\text{COP}_\text{c}}\)\end{tabular}}} & \(Q_{\text{cold}} \times \frac{\text{COP}_\text{R}}{\text{COP}_\text{c}}\)  & \(Q_{\text{cold}} \times \frac{\text{COP}_\text{R}}{\text{COP}_\text{c}}\) & \multicolumn{1}{c|}{\textbf{\begin{tabular}[c]{@{}c@{}} Output \\ \(Q_\text{cold}\)\\ (meV)\end{tabular}}} & \multicolumn{1}{c|}{\textbf{\begin{tabular}[c]{@{}c@{}} \(\frac{\text{COP}_\text{R}}{\text{COP}_\text{c}}\)\end{tabular}}} & \(Q_{\text{cold}} \times \frac{\text{COP}_\text{R}}{\text{COP}_\text{c}}\) \\ \hline
\multirow{3}{*}{\textbf{\(r_c\) = 3.5}}                                                          & \textbf{Bilayer}                                                                    & \multicolumn{1}{c|}{0.464}                                                                                 & \multicolumn{1}{c|}{\color[HTML]{3197FF} \textbf{0.800 }}                                                                    & 0.371                & \multicolumn{1}{c|}{0.464}                                                                                 & \multicolumn{1}{c|}{\color[HTML]{3197FF} \textbf{0.800}}                                                                    & 0.371                & 0.496             &\multicolumn{1}{c|} {--} & \multicolumn{1}{c|} {--} & \multicolumn{1}{c|} {--}   \\ \cline{2-12} 
                                                                                            & \textbf{TBG (\(0.7^o\))}                                                                  & \multicolumn{1}{c|}{1.171}                                                                                 & \multicolumn{1}{c|}{0.61}                                                                     & \color[HTML]{3197FF} \textbf{0.714}       & \multicolumn{1}{c|}{1.058}                                                                                 & \multicolumn{1}{c|}{0.475}                                                                    & \color[HTML]{3197FF} \textbf{0.503}       & 1.356        &\multicolumn{1}{c|} {--} & \multicolumn{1}{c|} {--} & \multicolumn{1}{c|} {--}        \\ \cline{2-12} 
                                                                                            & \textbf{MATBG (\(1.05^o\))}                                                               & \multicolumn{1}{c|}{\color[HTML]{3197FF} \textbf{1.72}}                                                                                  & \multicolumn{1}{c|}{0.243}                                                                    & 0.418                & \multicolumn{1}{c|}{\color[HTML]{3197FF} \textbf{1.692 }}                                                                                 & \multicolumn{1}{c|}{0.219}                                                                    & 0.371                & \color[HTML]{3197FF} \textbf{2.919}   &\multicolumn{1}{c|} {--} & \multicolumn{1}{c|} {--} & \multicolumn{1}{c|} {--}    \\ \hline
\multirow{3}{*}{\textbf{\(r_c\) = 4.5}}                                                          & \textbf{Bilayer}                                                                    & \multicolumn{1}{c|}{1.018}                                                                                 & \multicolumn{1}{c|}{\color[HTML]{3197FF} \textbf{0.571 }}                                                                    & 0.582                & \multicolumn{1}{c|}{1.018}                                                                                 & \multicolumn{1}{c|}{\color[HTML]{3197FF} \textbf{0.571}}                                                                    & 0.582                & 1.231               &\multicolumn{1}{c|} {--} & \multicolumn{1}{c|} {--} & \multicolumn{1}{c|} {--}  \\ \cline{2-12} 
                                                                                            & \textbf{TBG (\(0.7^o\))}                                                                  & \multicolumn{1}{c|}{1.829}                                                                                 & \multicolumn{1}{c|}{0.439}                                                                    & \color[HTML]{3197FF} \textbf{0.802}       & \multicolumn{1}{c|}{1.722}                                                                                 & \multicolumn{1}{c|}{0.374}                                                                    & \color[HTML]{3197FF} \textbf{0.644}       & 2.337     &\multicolumn{1}{c|} {--} & \multicolumn{1}{c|} {--} & \multicolumn{1}{c|} {--}           \\ \cline{2-12} 
                                                                                            & \textbf{MATBG (\(1.05^o\))}                                                               & \multicolumn{1}{c|}{\color[HTML]{3197FF} \textbf{2.003}}                                                                                 & \multicolumn{1}{c|}{0.171}                                                                    & 0.343                & \multicolumn{1}{c|}{\color[HTML]{3197FF} \textbf{1.988}}                                                                                 & \multicolumn{1}{c|}{0.153}                                                                    & 0.304                & \color[HTML]{3197FF} \textbf{3.941}    &\multicolumn{1}{c|} {--} & \multicolumn{1}{c|} {--} & \multicolumn{1}{c|} {--}   \\ \hline
\multirow{3}{*}{\textbf{\(r_c\) = 5.5}}                                                          & \textbf{Bilayer}                                                                    & \multicolumn{1}{c|}{1.316}                                                                                 & \multicolumn{1}{c|}{\color[HTML]{3197FF} \textbf{0.444}}                                                                    & 0.584                & \multicolumn{1}{c|}{1.316}                                                                                 & \multicolumn{1}{c|}{\color[HTML]{3197FF} \textbf{0.444}}                                                                    & 0.584                & 1.497             &\multicolumn{1}{c|} {--} & \multicolumn{1}{c|} {--} & \multicolumn{1}{c|} {--}   \\ \cline{2-12} 
                                                                                            & \textbf{TBG (\(0.7^o\))}                                                                  & \multicolumn{1}{c|}{\color[HTML]{3197FF} \textbf{2.282 }}                                                                                 & \multicolumn{1}{c|}{0.34}                                                                     & \color[HTML]{3197FF} \textbf{0.777}       & \multicolumn{1}{c|}{\color[HTML]{3197FF} \textbf{2.204}}                                                                                 & \multicolumn{1}{c|}{0.306}                                                                    & \color[HTML]{3197FF} \textbf{0.675}       & 3.151            &\multicolumn{1}{c|} {--} & \multicolumn{1}{c|} {--} & \multicolumn{1}{c|} {--}    \\ \cline{2-12} 
                                                                                            & \textbf{MATBG (\(1.05^o\))}                                                               & \multicolumn{1}{c|}{2.158}                                                                                 & \multicolumn{1}{c|}{0.133}                                                                    & 0.287                & \multicolumn{1}{c|}{2.147}                                                                                 & \multicolumn{1}{c|}{0.118}                                                                    & 0.253                & \color[HTML]{3197FF} \textbf{4.731}    &\multicolumn{1}{c|} {--} & \multicolumn{1}{c|} {--} & \multicolumn{1}{c|} {--}   \\ \hline
\multirow{1}{*}{\textbf{\(r_c\) = 0.3}}                                                          & \textbf{MATBG (\(1.05^o\))}                                                                    & \multicolumn{1}{c|}{--}                                                                                 & \multicolumn{1}{c|}{--}                                                                    & --                & \multicolumn{1}{c|}{--}                                                                                 & \multicolumn{1}{c|}{--}                                                                    & --               & --               
                                                                                              & \multicolumn{1}{c|}{0.901} & \multicolumn{1}{c|}{0.088} & \multicolumn{1}{c|}{0.079}   \\ \hline
\end{tabular}
\caption{COP, Refrigeration Output in (meV) and coefficient of merit (\(Q_\text{cold} \times \frac{\text{COP}_{\text{R}}}{\text{COP}_{\text{c}}}\)) for QOC (general and strict) and QCC for specific compression ratio values operating in \textbf{Refrigeration} regime. The COP is in units of Carnot COP (\(\text{COP}_c=0.50\)). Parameters : \( T_\text{h} = 150\,\mathrm{K} \), \( T_\text{c} = 50\,\mathrm{K} \), and \( B_1 = 1\,\mathrm{T} \).}
\label{tab:re_comparison}
\end{table*}

\small{
\begin{table*}[!htbp]
\begin{tabular}{|c|c|ccc|ccc|ccc|}
\hline
\multirow{2}{*}{\textbf{\begin{tabular}[c]{@{}c@{}}Compression\\ Ratio Value\end{tabular}}} & \multirow{2}{*}{\textbf{\begin{tabular}[c]{@{}c@{}}Graphene\\ System\end{tabular}}} & \multicolumn{3}{c|}{\textbf{\begin{tabular}[c]{@{}c@{}}General QOC\end{tabular}}}                                                                          & \multicolumn{3}{c|}{\textbf{\begin{tabular}[c]{@{}c@{}}Strict QOC\end{tabular}}}                                                                           & \multicolumn{3}{c|}{\textbf{QSC}}                                                                                                                             \\ \cline{3-11} 
                                                                                            &                                                                                     & \multicolumn{1}{c|}{\textbf{\begin{tabular}[c]{@{}c@{}}Output\\  \(Q_\text{cold}\) (meV)\end{tabular}}} & \multicolumn{1}{c|}{\textbf{\(\text{COP}_\text{CP}\)}} & {\footnotesize\( Q_\text{cold} \times \text{COP}_\text{CP} \)} & \multicolumn{1}{c|}{\textbf{\begin{tabular}[c]{@{}c@{}} Output\\  \(Q_\text{cold}\) (meV)\end{tabular}}} & \multicolumn{1}{c|}{\textbf{\(\text{COP}_\text{CP}\)}} & {\footnotesize\( Q_\text{cold} \times \text{COP}_\text{CP} \)} & \multicolumn{1}{c|}{\textbf{\begin{tabular}[c]{@{}c@{}}Output\\  \(Q_\text{cold}\) (meV)\end{tabular}}} & \multicolumn{1}{c|}{\textbf{\(\text{COP}_\text{CP}\)}} & {\footnotesize\( Q_\text{cold} \times \text{COP}_\text{CP} \)} \\ \hline
\multirow{5}{*}{\textbf{\(r_c\) = 0.65}}                                                         & \textbf{Monolayer}                                                                  & \multicolumn{1}{c|}{9.474}                                                                         & \multicolumn{1}{c|}{5.109}        & 48.404               & \multicolumn{1}{c|}{9.474}                                                                         & \multicolumn{1}{c|}{5.109}        & 48.404               & \multicolumn{1}{c|}{4.710}                                                                         & \multicolumn{1}{c|}{3.536}        & 16.654               \\ \cline{2-11} 
                                                                                            & \textbf{Bilayer}                                                                    & \multicolumn{1}{c|}{14.892}                                                                        & \multicolumn{1}{c|}{2.832}        & 42.169               & \multicolumn{1}{c|}{14.892}                                                                        & \multicolumn{1}{c|}{2.832}        & 42.169               & \multicolumn{1}{c|}{6.351}                                                                         & \multicolumn{1}{c|}{1.786}        & 11.340               \\ \cline{2-11} 
                                                                                            & \textbf{TBG (\( 0.7^o\))}                                                                  & \multicolumn{1}{c|}{14.774}                                                                        & \multicolumn{1}{c|}{2.415}        & 35.68                & \multicolumn{1}{c|}{14.844}                                                                        & \multicolumn{1}{c|}{2.385}        & 35.398               & \multicolumn{1}{c|}{5.202}                                                                         & \multicolumn{1}{c|}{1.344}        & 6.994                \\ \cline{2-11} 
                                                                                            & \textbf{MATBG (\( 1.05^o\))}                                                               & \multicolumn{1}{c|}{12.453}                                                                        & \multicolumn{1}{c|}{1.800}        & 22.422               & \multicolumn{1}{c|}{12.736}                                                                        & \multicolumn{1}{c|}{1.762}        & 22.437               & \multicolumn{1}{c|}{2.376}                                                                         & \multicolumn{1}{c|}{0.640}        & 1.522                \\ \cline{2-11} 
                                                                                            & \textbf{TBG (\( 3.0^o\))}                                                                  & \multicolumn{1}{c|}{\color[HTML]{3197FF} \textbf{16.374}}                                                                        & \multicolumn{1}{c|}{\color[HTML]{3197FF} \textbf{5.154}}        & \color[HTML]{3197FF} \textbf{84.396}      & \multicolumn{1}{c|}{\color[HTML]{3197FF} \textbf{16.376}}                                                                        & \multicolumn{1}{c|}{\color[HTML]{3197FF} \textbf{5.152}}        & \color[HTML]{3197FF} \textbf{84.365}      & \multicolumn{1}{c|}{\color[HTML]{3197FF} \textbf{9.242}}                                                                         & \multicolumn{1}{c|}{\color[HTML]{3197FF} \textbf{3.804}}        & \color[HTML]{3197FF} \textbf{35.159}      \\ \hline
\multirow{5}{*}{\textbf{\(r_c\) = 0.75}}                                                         & \textbf{Monolayer}                                                                  & \multicolumn{1}{c|}{7.638}                                                                         & \multicolumn{1}{c|}{7.37}         & 56.291               & \multicolumn{1}{c|}{7.638}                                                                         & \multicolumn{1}{c|}{7.37}         & 56.291               & \multicolumn{1}{c|}{4.733}                                                                         & \multicolumn{1}{c|}{5.752}        & 27.222               \\ \cline{2-11} 
                                                                                            & \textbf{Bilayer}                                                                    & \multicolumn{1}{c|}{12.238}                                                                        & \multicolumn{1}{c|}{3.953}        & 48.38                & \multicolumn{1}{c|}{12.238}                                                                        & \multicolumn{1}{c|}{3.953}        & 48.38                & \multicolumn{1}{c|}{6.898}                                                                         & \multicolumn{1}{c|}{2.92}         & 20.143               \\ \cline{2-11} 
                                                                                            & \textbf{TBG (\( 0.7^o\))}                                                                  & \multicolumn{1}{c|}{11.81}                                                                         & \multicolumn{1}{c|}{3.306}        & 39.05                & \multicolumn{1}{c|}{11.837}                                                                        & \multicolumn{1}{c|}{3.272}        & 38.728               & \multicolumn{1}{c|}{5.858}                                                                         & \multicolumn{1}{c|}{2.259}        & 13.232               \\ \cline{2-11} 
                                                                                            & \textbf{MATBG (\( 1.05^o\))}                                                               & \multicolumn{1}{c|}{9.108}                                                                         & \multicolumn{1}{c|}{2.332}        & 21.242               & \multicolumn{1}{c|}{9.204}                                                                         & \multicolumn{1}{c|}{2.289}        & 21.065               & \multicolumn{1}{c|}{2.988}                                                                         & \multicolumn{1}{c|}{1.202}        & 3.592                \\ \cline{2-11} 
                                                                                            & \textbf{TBG (\( 3.0^o\))}                                                                  & \multicolumn{1}{c|}{\color[HTML]{3197FF} \textbf{13.808}}                                                                        & \multicolumn{1}{c|}{\color[HTML]{3197FF} \textbf{7.421}}        & \color[HTML]{3197FF} \textbf{102.465}     & \multicolumn{1}{c|}{\color[HTML]{3197FF} \textbf{13.809}}                                                                        & \multicolumn{1}{c|}{\color[HTML]{3197FF} \textbf{7.418}}        & \color[HTML]{3197FF} \textbf{102.439}     & \multicolumn{1}{c|}{\color[HTML]{3197FF} \textbf{9.319}}                                                                         & \multicolumn{1}{c|}{\color[HTML]{3197FF} \textbf{6.043}}        & \color[HTML]{3197FF} \textbf{56.316}      \\ \hline
\multirow{5}{*}{\textbf{\(r_c\) = 0.85}}                                                         & \textbf{Monolayer}                                                                  & \multicolumn{1}{c|}{6.266}                                                                         & \multicolumn{1}{c|}{12.572}       & 78.776               & \multicolumn{1}{c|}{6.266}                                                                         & \multicolumn{1}{c|}{12.572}       & 78.776               & \multicolumn{1}{c|}{4.747}                                                                         & \multicolumn{1}{c|}{10.911}       & 51.794               \\ \cline{2-11} 
                                                                                            & \textbf{Bilayer}                                                                    & \multicolumn{1}{c|}{10.22}                                                                         & \multicolumn{1}{c|}{6.547}        & 66.904               & \multicolumn{1}{c|}{10.22}                                                                         & \multicolumn{1}{c|}{6.547}        & 66.904               & \multicolumn{1}{c|}{7.359}                                                                         & \multicolumn{1}{c|}{5.525}        & 40.657               \\ \cline{2-11} 
                                                                                            & \textbf{TBG (\( 0.7^o\))}                                                                  & \multicolumn{1}{c|}{9.607}                                                                         & \multicolumn{1}{c|}{5.374}        & 51.625               & \multicolumn{1}{c|}{9.615}                                                                         & \multicolumn{1}{c|}{5.333}        & 51.277               & \multicolumn{1}{c|}{6.426}                                                                         & \multicolumn{1}{c|}{4.347}        & 27.929               \\ \cline{2-11} 
                                                                                            & \textbf{MATBG (\( 1.05^o\))}                                                               & \multicolumn{1}{c|}{6.715}                                                                         & \multicolumn{1}{c|}{3.587}        & 24.085               & \multicolumn{1}{c|}{6.739}                                                                         & \multicolumn{1}{c|}{3.538}        & 23.844               & \multicolumn{1}{c|}{3.521}                                                                         & \multicolumn{1}{c|}{2.485}        & 8.752                \\ \cline{2-11} 
                                                                                            & \textbf{TBG (\( 3.0^o\))}                                                                  & \multicolumn{1}{c|}{\color[HTML]{3197FF} \textbf{\color[HTML]{3197FF} \textbf{11.788}}}                                                                        & \multicolumn{1}{c|}{\color[HTML]{3197FF} \textbf{12.609}}       & \color[HTML]{3197FF} \textbf{148.631}     & \multicolumn{1}{c|}{\color[HTML]{3197FF} \textbf{11.788}}                                                                        & \multicolumn{1}{c|}{\color[HTML]{3197FF} \textbf{12.607}}       & \color[HTML]{3197FF} \textbf{148.609}     & \multicolumn{1}{c|}{\color[HTML]{3197FF} \textbf{9.37}}                                                                          & \multicolumn{1}{c|}{\color[HTML]{3197FF} \textbf{11.205}}       & \color[HTML]{3197FF} \textbf{104.996}     \\ \hline
\end{tabular}
\caption{COP, Heat Output in (meV) and coefficient of merit (\(Q_\text{cold} \times \frac{\text{COP}_{\text{CP}}}{\text{COP}_{\text{c}}}\)) for QOC (general and strict) and QSC for specific compression ratio values operating in \textbf{Cold Pump} regime. Parameters : \( T_\text{h} = 150\,\mathrm{K} \), \( T_\text{c} = 50\,\mathrm{K} \), and \( B_1 = 1\,\mathrm{T} \).}
\label{tab:cp_comparison}
\end{table*}
}

\begin{table*}[!htbp]
\begin{tabular}{|c|c|cccc|cccc|}
\hline
\multirow{2}{*}{\textbf{\begin{tabular}[c]{@{}c@{}}Compression\\ Ratio Value\end{tabular}}} & \multirow{2}{*}{\textbf{\begin{tabular}[c]{@{}c@{}}Graphene\\ System\end{tabular}}} & \multicolumn{4}{c|}{\textbf{QSC}}                                                                                                                                                                                                                                                                                                                         & \multicolumn{4}{c|}{\textbf{\begin{tabular}[c]{@{}c@{}}Strict \\ QOC\end{tabular}}}                                                                                                                                                                                                                                                                       \\ \cline{3-10} 
                                                                                            &                                                                                     & \multicolumn{1}{c|}{\textbf{\begin{tabular}[c]{@{}c@{}}Output\\ |\(Q_{\text{hot}}\)| (meV)\end{tabular}}} & \multicolumn{1}{c|}{\textbf{\begin{tabular}[c]{@{}c@{}}Output\\ |\(Q_{\text{cold}}\)| (meV)\end{tabular}}} & \multicolumn{1}{c|}{\textbf{\begin{tabular}[c]{@{}c@{}}Input\\ |W| (meV)\end{tabular}}} & \textbf{\begin{tabular}[c]{@{}c@{}}\(\Delta S\)\\ (meV/K)\end{tabular}} & \multicolumn{1}{c|}{\textbf{\begin{tabular}[c]{@{}c@{}}Output\\ |\(Q_{\text{hot}}\)| (meV)\end{tabular}}} & \multicolumn{1}{c|}{\textbf{\begin{tabular}[c]{@{}c@{}}Output\\ |\(Q_{\text{cold}}\)| (meV)\end{tabular}}} & \multicolumn{1}{c|}{\textbf{\begin{tabular}[c]{@{}c@{}}Input\\ |W| (meV)\end{tabular}}} & \textbf{\begin{tabular}[c]{@{}c@{}}\(\Delta S\)\\ (meV/K)\end{tabular}} \\ \hline
\multirow{4}{*}{\textbf{\(r_c\) = 0.32}}                                                         & \textbf{Monolayer}                                                                  & \multicolumn{1}{c|}{0.442}                                                                  & \multicolumn{1}{c|}{4.417}                                                                   & \multicolumn{1}{c|}{4.859}                                                              & \multicolumn{1}{c|}{\color[HTML]{3197FF} \textbf{0.091}}                                                              & \multicolumn{1}{c|}{--}                                                                     & \multicolumn{1}{c|}{--}                                                                      & \multicolumn{1}{c|}{--}                                                                 & --                                                                 \\ \cline{2-10} 
                                                                                            & \textbf{Bilayer}                                                                    & \multicolumn{1}{c|}{5.92}                                                                   & \multicolumn{1}{c|}{3.524}                                                                   & \multicolumn{1}{c|}{9.444}                                                              & 0.11                                                               & \multicolumn{1}{c|}{--}                                                                     & \multicolumn{1}{c|}{--}                                                                      & \multicolumn{1}{c|}{--}                                                                 & --                                                                 \\ \cline{2-10} 
                                                                                            & \textbf{TBG (\(0.7^o\))}                                                                  & \multicolumn{1}{c|}{7.993}                                                                  & \multicolumn{1}{c|}{2.012}                                                                   & \multicolumn{1}{c|}{10.005}                                                             & 0.094                                                              & \multicolumn{1}{c|}{--}                                                                     & \multicolumn{1}{c|}{--}                                                                      & \multicolumn{1}{c|}{--}                                                                 & --                                                                 \\ \cline{2-10} 
                                                                                            & \textbf{MATBG (\(1.05^o\))}                                                               & \multicolumn{1}{c|}{--}                                                                     & \multicolumn{1}{c|}{--}                                                                      & \multicolumn{1}{c|}{--}                                                                 & --                                                                 & \multicolumn{1}{c|}{--}                                                                     & \multicolumn{1}{c|}{--}                                                                      & \multicolumn{1}{c|}{--}                                                                 & --                                                                 \\ \hline
\multirow{4}{*}{\textbf{\(r_c\) = 0.5}}                                                          & \textbf{Monolayer}                                                                  & \multicolumn{1}{c|}{--}                                                                     & \multicolumn{1}{c|}{--}                                                                      & \multicolumn{1}{c|}{--}                                                                 & --                                                                 & \multicolumn{1}{c|}{--}                                                                     & \multicolumn{1}{c|}{--}                                                                      & \multicolumn{1}{c|}{--}                                                                 & --                                                                 \\ \cline{2-10} 
                                                                                            & \textbf{Bilayer}                                                                    & \multicolumn{1}{c|}{0.381}                                                                  & \multicolumn{1}{c|}{5.334}                                                                   & \multicolumn{1}{c|}{5.715}                                                              & 0.109                                                              & \multicolumn{1}{c|}{--}                                                                     & \multicolumn{1}{c|}{--}                                                                      & \multicolumn{1}{c|}{--}                                                                 & --                                                                 \\ \cline{2-10} 
                                                                                            & \textbf{TBG (\(0.7^o\))}                                                                  & \multicolumn{1}{c|}{2.12}                                                                   & \multicolumn{1}{c|}{4.024}                                                                   & \multicolumn{1}{c|}{6.144}                                                              & 0.095                                                              & \multicolumn{1}{c|}{--}                                                                     & \multicolumn{1}{c|}{--}                                                                      & \multicolumn{1}{c|}{--}                                                                 & --                                                                 \\ \cline{2-10} 
                                                                                            & \textbf{MATBG (\(1.05^o\))}                                                               & \multicolumn{1}{c|}{4.615}                                                                  & \multicolumn{1}{c|}{1.284}                                                                   & \multicolumn{1}{c|}{5.899}                                                              & \multicolumn{1}{c|}{\color[HTML]{3197FF} \textbf{0.056}}                                                              & \multicolumn{1}{c|}{--}                                                                     & \multicolumn{1}{c|}{--}                                                                      & \multicolumn{1}{c|}{--}                                                                 & --                                                                 \\ \hline
\multirow{2}{*}{\textbf{\(r_c\) = 1.73}}                                                         & \textbf{TBG (\(0.7^o\))}                                                                  & \multicolumn{1}{c|}{--}                                                                     & \multicolumn{1}{c|}{--}                                                                      & \multicolumn{1}{c|}{--}                                                                 & --                                                                 & \multicolumn{1}{c|}{0.125}                                                                  & \multicolumn{1}{c|}{0.043}                                                                   & \multicolumn{1}{c|}{0.168}                                                              & 0.002                                                              \\ \cline{2-10} 
                                                                                            & \textbf{MATBG (\(1.05^o\))}                                                               & \multicolumn{1}{c|}{--}                                                                     & \multicolumn{1}{c|}{--}                                                                      & \multicolumn{1}{c|}{--}                                                                 & --                                                                 & \multicolumn{1}{c|}{--}                                                                     & \multicolumn{1}{c|}{--}                                                                      & \multicolumn{1}{c|}{--}                                                                 & --                                                                 \\ \hline
\multirow{2}{*}{\textbf{\(r_c\) = 2.5}}                                                          & \textbf{TBG (\(0.7^o\))}                                                                  & \multicolumn{1}{c|}{--}                                                                     & \multicolumn{1}{c|}{--}                                                                      & \multicolumn{1}{c|}{--}                                                                 & --                                                                 & \multicolumn{1}{c|}{--}                                                                     & \multicolumn{1}{c|}{--}                                                                      & \multicolumn{1}{c|}{--}                                                                 & --                                                                 \\ \cline{2-10} 
                                                                                            & \textbf{MATBG (\(1.05^o\))}                                                               & \multicolumn{1}{c|}{--}                                                                     & \multicolumn{1}{c|}{--}                                                                      & \multicolumn{1}{c|}{--}                                                                 & --                                                                 & \multicolumn{1}{c|}{0.073}                                                                  & \multicolumn{1}{c|}{0.169}                                                                   & \multicolumn{1}{c|}{0.242}                                                              & 0.004                                                              \\ \hline
\end{tabular}
\caption{Heat Output into hot reservoir (\(Q_\text{hot}\)) in meV, Heat Output into cold reservoir (\(Q_\text{cold}\)) in meV , Input Work (\(W\)) in meV and Entropy produced (\(\Delta S\)) in meV/K for strict QOC and QSC for specific compression ratio values operating in \textbf{Joule Pump} regime. Parameters : \( T_\text{h} = 150\,\mathrm{K} \), \( T_\text{c} = 50\,\mathrm{K} \), and \( B_1 = 1\,\mathrm{T} \).}
\label{tab:jp_comparison}
\end{table*}
\normalsize

{Since QSC is not designed with adiabatic strokes, heat transfer occurs during individual strokes. The operational phase in which QSC operates, depends on the relative signs of these individual heat transfers. Fig.~\ref{fig19} depicts the operational phase transitions as a function of compression ratio across the graphene systems. The refrigeration phase is characterized by \(Q_{CD} < Q_{DA} < 0 < Q_{BC} < Q_{AB}\), the Joule pump by \(Q_{CD} < Q_{DA} < 0 < Q_{AB} < Q_{BC}\), the cold pump by \(Q_{DA} < Q_{CD} < 0 < Q_{AB} < Q_{BC}\), and the heat engine by \(Q_{DA}, Q_{AB} < 0 < Q_{CD}, Q_{BC}\). We observe that the heat transferred during the isochoric processes (\(Q_{BC}\) and \(Q_{DA}\)), obtained from Eqs.~\eqref{eq:sterlingheat2} and \eqref{eq:sterlingheat4}, satisfies \(Q_{BC} > 0\) and \(Q_{DA} < 0\) across graphene systems irrespective of the value of compression ratios. Meanwhile, the heat exchanged during the isothermal processes, obtained from Eq.~\eqref{eq:sterlingheat1} (\(Q_{AB}\)), always decreases with compression ratio, whereas Eq.~\eqref{eq:sterlingheat3} (\(Q_{CD}\)) increases with compression ratio for all graphene platforms. Since $Q_{\text{hot}}$ represents the heat exchanged during the hot isochoric ($Q_{BC}$) and hot isothermal ($Q_{CD}$) processes, and $Q_{\text{cold}}$ is the heat exchanged during the cold isochoric ($Q_{DA}$) and cold isothermal ($Q_{AB}$) processes, the relative magnitudes of these four contributions determine the operational phase of the cycle. The lower Landau-level energies in MATBG reduce the heat exchanged during the isochoric strokes (Eqs.~\eqref{eq:sterlingheat2} and~\eqref{eq:sterlingheat4}), leading to an additional refrigeration phase at very low compression ratios, as observed in Fig.~\ref{fig19}(c).}

\subsection{Comparative analysis of heat engine phase across different quantum thermodynamic cycles.}

Table~\ref{tab:he_comparison} compares the efficiency and work output for various quantum thermodynamic cycles operating in the heat engine regime across selected compression ratio values. The QOC yields nearly identical results under strict and general adiabatic conditions at low compression ratios. However, at higher compression ratios, the general adiabatic condition improves efficiency compared to the strict condition. MATBG consistently demonstrates the highest efficiency in both QOC variants and the QSC. 

In the QOC, this high efficiency is accompanied by a significantly reduced work output, which diminishes MATBG’s overall coefficient of merit (\(\text{W} \times \frac{\eta}{\eta_c}\)). Conversely, in the QSC, MATBG maintains high efficiency without compromising work output, making it the most favorable regime for operation. For the QCC, all systems inherently operate at Carnot efficiency~\((\eta_c)\), with MATBG exhibiting the lowest work output, making it the least favorable due to its reduced coefficient of merit.


\subsection{Comparative analysis of refrigeration phase across different quantum thermodynamic cycles.}

Table~\ref{tab:re_comparison} compares the performance of various quantum thermodynamic cycles in the refrigeration regime. For the QOC, both strict and general adiabatic implementations yield nearly identical results at low compression ratios. In both cases, increasing the compression ratio \(r_c\) reduces the COP and a corresponding increase in refrigeration output. Across these QOC configurations, MATBG consistently exhibits the lowest COP among the graphene systems. However, its refrigeration output is relatively high at lower compression ratios, suggesting effective refrigeration for low compression ratios. Bilayer graphene attains the highest COP in this regime, with a comparatively lower Refrigeration Output, and TBG (at \(\theta =0.7^o\)) attains the highest coefficient of merit (\( Q_\text{cold} \times \frac{\text{COP}_\text{R}}{\text{COP}_\text{c}}\)) across the refrigeration phase.

\begin{figure*}[!htbp]
    \centering
    \includegraphics[width=1\linewidth]{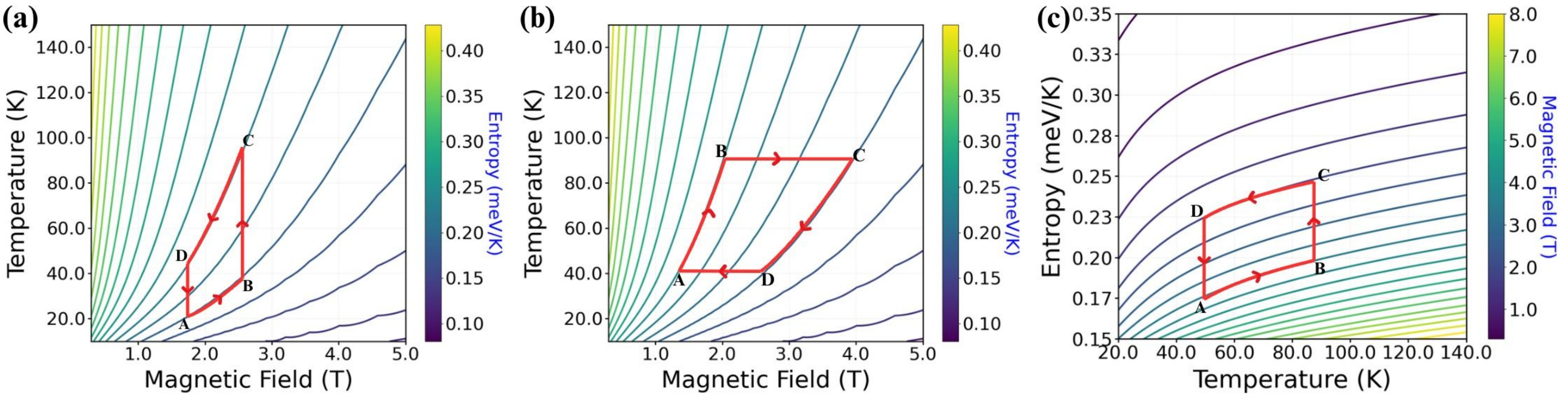}
    \caption{Isoentropic contour line on the temperature–magnetic field (T–B) plane in MATBG for (a) QOC, (b) QCC, and isoenergetic contour line on the entropy-temperature (S-T) plane in MATBG for (c) QSC}
    \label{fig20}
\end{figure*}

In the QCC, where all systems operate at the Carnot limit \(\text{COP}_c\). MATBG demonstrates the highest Refrigeration Output in this setting, highlighting its superior overall performance when operating reversibly in the QCC. The refrigeration phase is observed exclusively in MATBG for the chosen reservoir temperature and only at very low compression ratios (\(r_c\approx 0.3\)), with a low coefficient of performance compared to QOC and QCC. For \(r_c \ge 1\), Fig.~\ref{fig20} shows that, at fixed temperature, increasing \(B\) leads to a decrease in entropy. Consequently, from Eq.~\eqref{eq:sterlingheat1}, \(Q_\text{AB} \le 0\). Additionally, under all parameter choices, Eq.~\eqref{eq:sterlingheat4} yields \(Q_\text{DA} \le 0\), implying that \(Q_\text{cold} \le 0\). Therefore, the quantum Stirling cycle (QSC) does not support a refrigeration phase for \(r_c \ge 1\).

\vspace{-1.5em}
\subsection{Comparative analysis of cold pump phase across different quantum thermodynamic cycles.}

Table~\ref{tab:cp_comparison} presents a comparative analysis of the performance of various graphene systems operating in the cold pump regime across different quantum thermodynamic cycles. As the compression ratio increases, both the strict and general versions of the QOC exhibit a decrease in the heat output (\(|Q_\text{cold}|\)) to the cold reservoir and an increase in the COP. The QSC shows simultaneous increases in both heat output and COP with increasing compression ratio.

Across all cycles, MATBG consistently demonstrates the lowest COP and heat output to the cold reservoir, while TBG at larger twist angles exhibits best coefficient of merit (\( Q_\text{cold} \times \frac{\text{COP}_\text{R}}{\text{COP}_\text{c}}\)). The results from strict and general implementations of the QOC align closely, with only minor deviations observed at smaller twist angles. For a fixed compression ratio, the QSC yields both lower COP and reduced heat output compared to either QOC variant. Due to the reversibility condition for QCC (Eq.~\eqref{eq: QCC condition}), no cold pumping regime is observed at any value of the compression ratio.

\subsection{Comparative analysis of Joule pump phase across different quantum thermodynamic cycles.}

Table~\ref{tab:jp_comparison} presents a comparative analysis of strict QOC and QSC across various graphene based platforms. In systems operating under QSC protocols, the Joule pump phase is observed for all platforms except TBG at higher twist angles. Monolayer graphene enters the Joule pump regime only at compression ratios near \( r_c \approx 0.3 \), and exits this regime as \( r_c \) increases. MATBG consistently demonstrates the lowest entropy production within the Joule pump phase, outperforming other platforms in terms of reversibility. Under strict QOC operation, the Joule pump regime is observed only for small windows of \( r_c \) in TBG systems, while MATBG exhibits this phase at low \( r_c \) with notably reduced entropy production. In both protocols, TBG with higher twist angles fails to enter the Joule pump regime.

    
    


\section{ Experimental Realization and Conclusion}
\label{sec:conc}

Fig.~\ref{fig20} illustrates how quantum thermodynamic cycles can, in principle, be implemented in magic-angle twisted bilayer graphene (MATBG) through careful modulation of magnetic field and temperature. The feasibility of such implementations hinges on realizing each thermodynamic stroke individually: isochoric processes can be achieved by maintaining a constant magnetic field while thermally coupling the system to a reservoir; isothermal strokes require quasi-static modulation of the magnetic field while in contact with a thermal bath; and adiabatic strokes demand slow, coherent evolution of the system’s Hamiltonian. These protocols, while conceptually straightforward, necessitate precise control over local parameters such as twist angle, magnetic flux density, and gate-defined electrostatic potentials. Encouragingly, recent advances in thermodynamic characterization, such as entropy mapping via single-electron transistors and scanning thermoelectric probes, have demonstrated the capacity to extract local temperature and entropy in moiré superlattices~\cite{rozen2021entropic, abualnaja2023entropy, pyurbeeva2021entropy}. These techniques represent a crucial experimental foundation for realizing full quantum thermodynamic cycles in MATBG and related strongly correlated two-dimensional materials.

In this article, we have presented a comprehensive quantum thermodynamic analysis of graphene based systems subjected to tunable magnetic fields, spanning monolayer graphene, AB-stacked bilayer graphene, twisted bilayer graphene (TBG), and magic-angle TBG (MATBG). By systematically examining their behavior across the Quantum Otto, Carnot, and Stirling cycles, we identified distinct operational modes, including heat engine, refrigeration, cold pump, and Joule pump, and assessed their performance as a function of compression ratio and twist angle. Notably, MATBG consistently enters the heat engine regime across all cycles. It achieves a particularly high coefficient of merit in the Stirling cycle and high efficiency in the Otto cycle, albeit with modest work output. In refrigeration and cold pump modes, MATBG’s performance is more subdued under QOC and QSC, but it surpasses other graphene systems in the QCC configuration, achieving higher cooling power at Carnot-limited coefficients of performance. Furthermore, the emergence of a Joule pump mode in QSC and strict QOC settings underscores MATBG’s unique thermodynamic versatility.

Looking forward, MATBG and similar moiré systems offer an exciting testbed for quantum thermodynamics in regimes where flat-band physics intertwines with quantum coherence. Future work could explore Magic angle twisted trilayer graphene (MATTG),  in modifying thermodynamic response functions in such systems. Experimentally, the integration of real-time calorimetry, magneto-transport, and entropy-sensitive probes with cycle protocols could enable direct observation of quantum work extraction and heat flow in moiré platforms. The ability to engineer and optimize quantum thermal machines using designer band structures opens new avenues for quantum energy harvesting, low-temperature cooling, and information-to-energy conversion at the nanoscale.

\bibliographystyle{unsrt} 

\bibliography{ref}

\clearpage
\appendix

\subsection*{Appendix A: Existence of different operational phases for quantum thermodynamic cycles with two reservoirs.}

\renewcommand{\theequation}{A\arabic{equation}}  
\setcounter{equation}{0}

The first law of thermodynamics states that the total energy of a system is conserved \citep{Quan2007,lucio2025innovative}. It is expressed as
\begin{equation}
    \Delta U = Q - W,
\end{equation}
\( \Delta U \) denotes the change in internal energy of the system, due to the heat absorbed \( Q \) and the work performed \( W \). For a thermodynamic cycle working between a hot reservoir at temperature \( T_\text{h} \) and a cold reservoir at temperature \( T_\text{c} \), exchanging heat \( Q_\text{hot} \) and \( Q_\text{cold} \), the first law reduces to,

\begin{equation}
    W = Q_\text{hot} + Q_\text{cold},
    \label{eq:appendixa firstlaw}
\end{equation}
since the internal energy returns to its initial value over a full cycle (\( \Delta U = 0 \)). The second law of thermodynamics states that the total entropy of the universe must increase or remain constant in a thermodynamic process \citep{landi2021irreversible,lucio2025innovative}. This is given by,
\begin{equation}
    \Delta S_\text{uni} = \Delta S_\text{sys} + \Delta S_\text{reservoir} \ge 0.
\end{equation}

For a cyclic process, the entropy change of the system is zero (\( \Delta S_\text{sys} = 0 \)), so the entropy change of the reservoirs must satisfy
\begin{equation}
    \Delta S_\text{reservoir} = \frac{Q_\text{hot}^{\text{reservoir}}}{T_\text{h}} + \frac{Q_\text{cold}^{\text{reservoir}}}{T_\text{c}} \ge 0.
\end{equation}
Since the reservoir releases the heat absorbed by the system, we have \( Q_\text{hot}^{\text{reservoir}} = -Q_\text{hot} \) and \( Q_\text{cold}^{\text{reservoir}} = -Q_\text{cold} \), leading to

\begin{equation}
    \frac{Q_\text{hot}}{T_\text{h}} + \frac{Q_\text{cold}}{T_\text{c}} \le 0.
    \label{eq:appnedixa:2ndlaw}
\end{equation}

We define the dimensionless parameters : \(\alpha = \frac{T_\text{h}}{T_\text{c}}\) and \(\beta = \frac{|Q_\text{hot}|}{|Q_\text{cold}|}\). Based on the signs of \(W\), \(Q_\text{hot}\), and \(Q_\text{cold}\), eight distinct thermodynamic configurations arise.

The combinations \(W \ge 0,\ Q_\text{hot} \le 0,\ Q_\text{cold} \le 0\) and \(W \le 0,\ Q_\text{hot} \ge 0,\ Q_\text{cold} \ge 0\) violate the first law (Eq.~\eqref{eq:appendixa firstlaw}), while \(W \ge 0,\ Q_\text{hot} \ge 0,\ Q_\text{cold} \ge 0\) violates the second law (Eq.~\eqref{eq:appnedixa:2ndlaw}). The case \(W \ge 0,\ Q_\text{hot} \le 0,\ Q_\text{cold} \ge 0\) leads to incompatible conditions, the first law requires \(\beta \le 1\), whereas the second law demands \(\beta \ge \alpha \ge 1\). Since these conditions cannot be satisfied simultaneously, this regime is thermodynamically forbidden as seen in Fig.~\ref{fig2}.

The four remaining configurations are physically viable. In the case of a heat engine (\(W \ge 0,\ Q_\text{hot} \ge 0,\ Q_\text{cold} \le 0\)), the first law implies \(\beta \ge 1\), while the second law requires \(\beta \le \alpha\), meaning this phase exists in the region \(1 \le \beta \le \alpha\). A cold pump (\(W \le 0,\ Q_\text{hot} \ge 0,\ Q_\text{cold} \le 0\)) satisfies both \(\beta \le 1\) and \(\beta \le \alpha\), thus existing in the domain \(0 \le \beta \le 1 \le \alpha\). For a refrigerator (\(W \le 0,\ Q_\text{hot} \le 0,\ Q_\text{cold} \ge 0\)), we find \(\beta \ge 1\) from the first law and \(\beta \ge \alpha \ge 1\) from the second, hence the region \(1 \le \alpha \le \beta < \infty\) is valid. It should be noted that as \(\beta \rightarrow \infty\), \(\text{COP}_R\) tends to zero \citep{lucio2025innovative}. Lastly, the Joule pump (\(W \le 0,\ Q_\text{hot} \le 0,\ Q_\text{cold} \le 0\)) imposes no constraint from the first law, allowing \(0\le\beta \le \infty\), and the second law enforces \(\beta \ge -\alpha\), which is trivially satisfied since \(\beta \ge 0\). Thus, this phase is also thermodynamically permitted.

\begin{widetext}

\begin{table}[!htbp]
\begin{tabular}{|c|c|c|c|c|}
\hline
\textbf{\(Q_{\text{hot}}\)}      & \textbf{\(Q_{\text{cold}}\)}       & \textbf{W}      & \textbf{Operation Phase}                     & \textbf{Constrains due to Laws of Thermodynamics}                                  \\ \hline
\textgreater{}0 & \textgreater{}0 & \textgreater{}0 & Violates \(II^{nd}\) Law                               & ---                                                              \\ \hline
\textgreater{}0 & \textgreater{}0 & \textless{}0    & Violates \(I^{st}\) and \(II^{nd}\) Law                         & ---                                                              \\ \hline
\textgreater{}0 & \textless{}0    & \textgreater{}0 & Heat Engine                                   & \begin{tabular}[c]{@{}c@{}}\(\beta\ge 1\) (First Law)\\ \(\beta \le \alpha\) (Second Law) \end{tabular}        \\ \hline
\textgreater{}0 & \textless{}0    & \textless{}0    & Cold Pump                                     & \begin{tabular}[c]{@{}c@{}}\(0 \le\beta\le 1\) (First Law)\\ \(\beta \le \alpha\) (Second Law)\end{tabular}        \\ \hline
\textless{}0    & \textgreater{}0 & \textgreater{}0 & \(I^{st}\) and \(II^{nd}\) law give non-intersecting constrains & \begin{tabular}[c]{@{}c@{}}\(\beta\le 1\)  (First Law) \\ \(\beta \ge \alpha \ge 1 \)  (Second Law)\end{tabular}      \\ \hline
\textless{}0    & \textgreater{}0 & \textless{}0    & Refrigerator                                  & \begin{tabular}[c]{@{}c@{}}\(\beta\ge 1\) (First Law)\\ \(\beta \ge \alpha\ge 1 \) (Second Law)\end{tabular}    \\ \hline
\textless{}0    & \textless{}0    & \textgreater{}0 & Violates \(I^{st}\) law                                & ---                                                              \\ \hline
\textless{}0    & \textless{}0    & \textless{}0    & Joule Pump                                    & \begin{tabular}[c]{@{}c@{}}\(0\le \beta \le \infty\)  (First Law)\\ \(\beta \ge-\alpha\) (Second Law)\end{tabular} \\ \hline
\end{tabular}
\caption{Different possible operational phases and their constraints given by the laws of thermodynamics.}
\end{table}
\end{widetext}

\subsection*{{Appendix B: Landau Level splitting and Occupation Probability in different graphene systems}}

\renewcommand{\theequation}{B\arabic{equation}}  
\setcounter{equation}{0}

\begin{figure}
    \centering
    \includegraphics[width=0.95\linewidth]{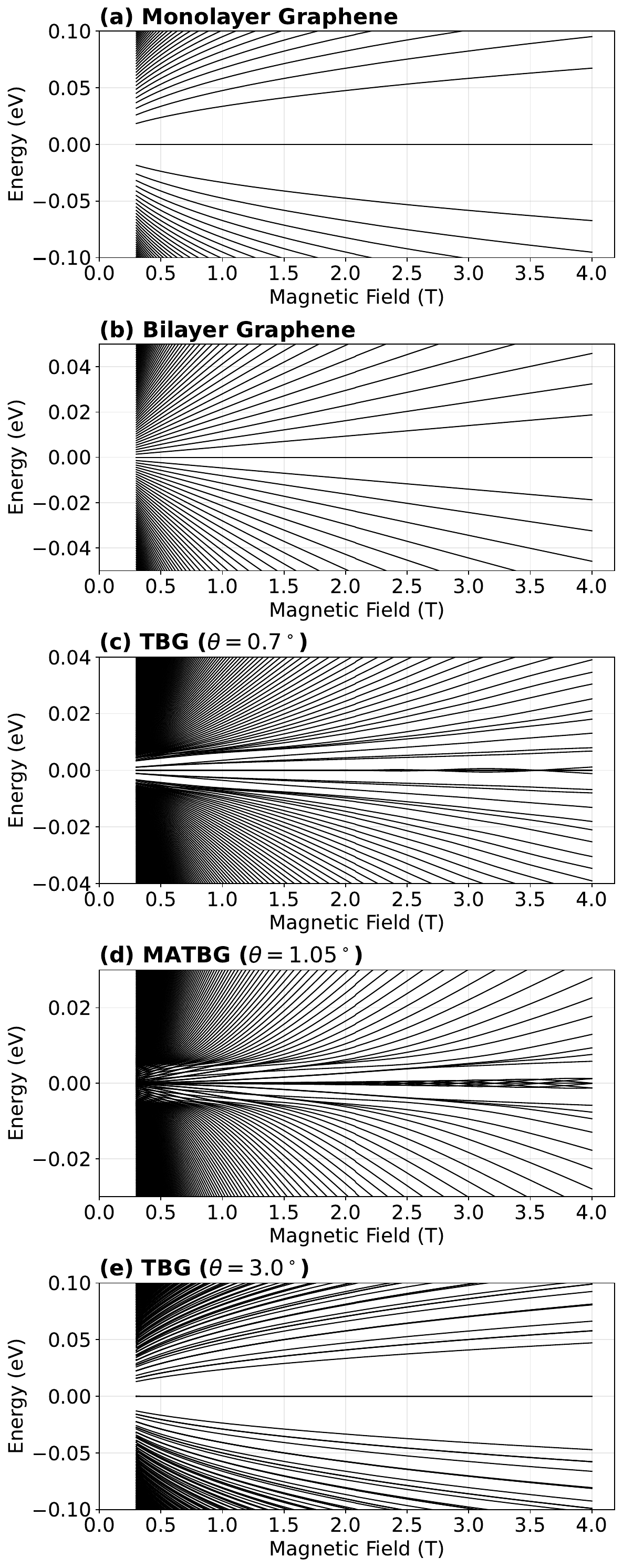}
    \caption{{Landau level splitting in various graphene based systems as a function of magnetic field.}}
    \label{Fig21}
\end{figure}

\begin{figure}
    \centering
    \includegraphics[width=0.95\linewidth]{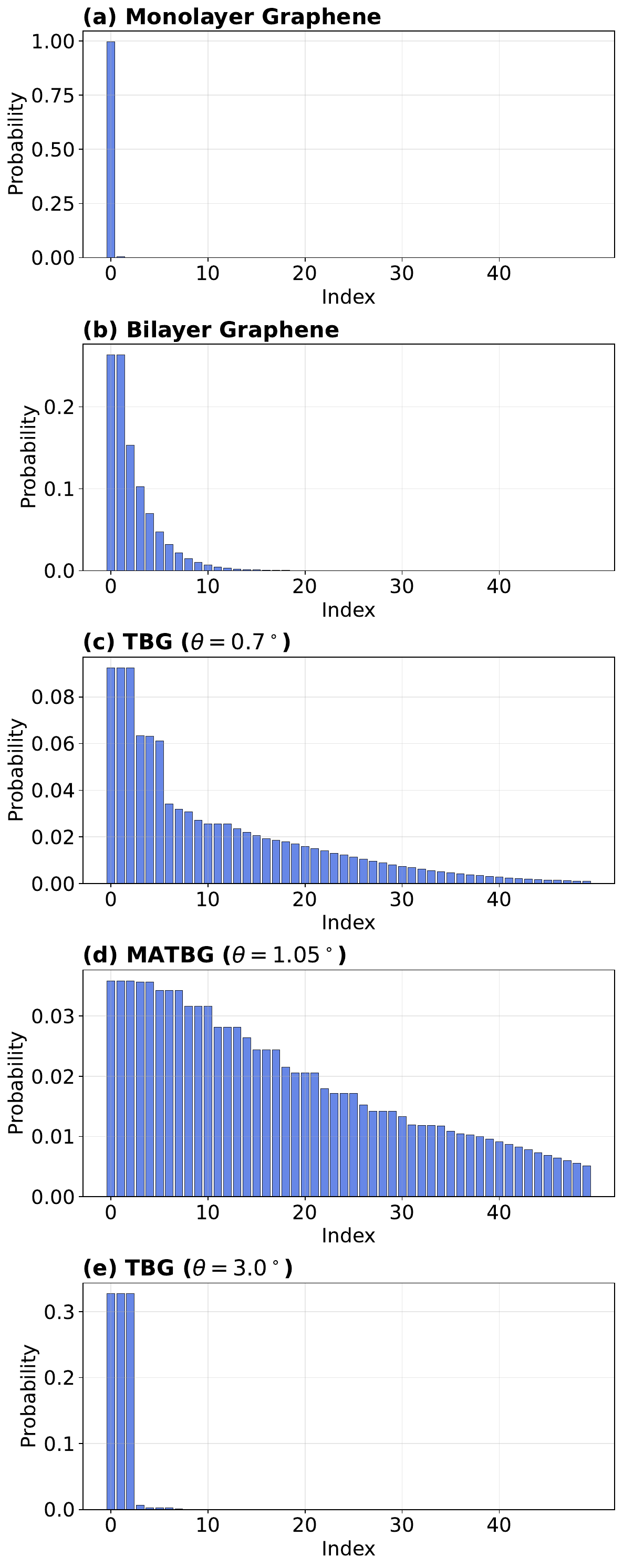}
    \caption{{Occupation probabilities of the lower Landau levels in graphene based systems at thermal equilibrium, with temperature $T = 100$ K and magnetic field $B = 1$ T.}}
    \label{Fig22}
\end{figure}

{
Fig.~\ref{Fig21} depicts the Landau level splitting for different graphene based systems. For larger twist angles, the TBG decouples and the Landau levels resemble those of monolayer graphene \citep{python2019}. At the magic angle, the Landau levels become tightly packed due to the flat-band dispersion. This allows electrons to occupy higher Landau levels, resulting in significant occupation probabilities across many levels, as seen in Fig.~\ref{Fig22}(d). 
}


\subsection*{Appendix C: Condition for equivalence of strict and general adiabatic conditions}
\renewcommand{\theequation}{C\arabic{equation}}  
\setcounter{equation}{0}

Consider an adiabatic stroke that evolves the system from an initial state defined by temperature \(T_{\alpha}\) and magnetic field \(B_{\alpha}\), to a final state with temperature \(T_{\beta}\) and magnetic field \(B_{\beta}\). The strict adiabatic condition requires that the occupation probabilities remain unchanged for all Landau levels,

\vspace{-1em}
\begin{equation}
    p_n(B_{\alpha}, T_{\alpha}) = p_n(B_{\beta}, T_{\beta}), \quad \forall n.
\end{equation}
\vspace{-1.5em}

Here, \(n\) denotes the Landau level index. The general adiabatic condition only demands that the von-Neumann entropy remains constant,

\vspace{-0.9em}
\begin{equation}
    S(B_{\alpha}, T_{\alpha}) = S(B_{\beta}, T_{\beta}),
    \label{eq:appendix_c_general}
\end{equation}
where \(S \) is given by Eq.\eqref{eq:von_neumann_entropy}. The strict condition implies the general one. We show that the converse holds under the assumption that all energy levels scale by a common factor \citep{singh2021magic},

\vspace{-1.5em}
\begin{equation}
    E_n(B_{\beta}) = \lambda E_n(B_{\alpha}), \quad \forall n.
    \label{eq: appendix energy scale}
\end{equation}
\vspace{-1.5em}



For initial and final states characterized by well-defined temperatures, with inverse temperatures related by \(\beta_1 = \lambda \beta_2\), the following condition holds \citep{singh2021magic},

\vspace{-1.5em}

\begin{align}
&\frac{p_n(B_{\alpha}, T_{\alpha})}{p_m(B_{\alpha}, T_{\alpha})} = \frac{p_n(B_{\beta}, T_{\beta})}{p_m(B_{\beta}, T_{\beta})} \nonumber \
\Rightarrow\quad \\&\frac{p_n(B_{\beta}, T_{\beta})}{p_n(B_{\alpha}, T_{\alpha})} = \frac{p_m(B_{\beta}, T_{\beta})}{p_m(B_{\alpha}, T_{\alpha})} = \phi, \quad \forall n, m.
\label{eq:phi_scaling_step2}
\end{align}

Substituting this into the entropy equality from Eq.~\eqref{eq:appendix_c_general} and using the von Neumann formula,
\begin{equation}
\begin{aligned}
S(B_{\alpha}, T_{\alpha})
&= \sum_n p_n(B_{\beta}, T_{\beta}) \ln p_n(B_{\beta}, T_{\beta}) \\
&= \sum_n \phi\, p_n(B_{\alpha}, T_{\alpha}) \ln(\phi\, p_n(B_{\alpha}, T_{\alpha})) \\
&= \phi \sum_n p_n(B_{\alpha}, T_{\alpha}) \ln p_n(B_{\alpha}, T_{\alpha}) + \phi \ln \phi
\end{aligned}
\label{eq:appendix_entropy_scaled}
\end{equation}

Hence, we obtain,
\begin{equation}
    (\phi - 1) S(B_{\alpha}, T_{\alpha}) = \phi \ln \phi.
\end{equation}

Since \(\phi\) must be independent of state variables and this equation must hold for arbitrary entropy \(S\), the only consistent solution is \(\phi = 1\). Therefore,
\begin{equation}
    p_n(B_{\alpha}, T_{\alpha}) = p_n(B_{\beta}, T_{\beta}), \quad \forall n,
\end{equation}
This confirms that the strict condition is satisfied. Monolayer and bilayer graphene obey the scaling condition in Eq.~\eqref{eq: appendix energy scale}, resulting in identical performance under strict and general adiabatic implementations of the QOC. TBG at all twist angles does not satisfy this condition, leading to observable differences between the two regimes.

\subsection*{Appendix D: Heat Exchanged during Isothermal Process}
\renewcommand{\theequation}{D\arabic{equation}}  
\setcounter{equation}{0}

For an isothermal process, where the system remains in contact with a thermal reservoir at temperature \(T\), we vary the external magnetic field from \(B_{\alpha}\) to \(B_{\beta}\) \citep{esposito2010entropy}. The expression for the heat exchanged is obtained from Eq.~\eqref{eq: Heat_work},
\begin{equation}
    Q = \sum_n \int_{B_{\alpha}}^{B_{\beta}} E_n(B) \frac{\partial p_n(B,T)}{\partial B} \, dB.
    \label{eq:appendix D Q}
\end{equation}

Taking the partial derivative with respect to \(B\) of the von Neumann entropy expression (see Eq.~\eqref{eq:von_neumann_entropy}) gives,

\begin{equation}
\begin{aligned}
    \frac{\partial S(B,T)}{\partial B}
    &= -k_B \sum_n \left[ \frac{\partial p_n(B,T)}{\partial B} \ln p_n(B,T) 
    + \frac{\partial p_n(B,T)}{\partial B} \right] \\
    &= -k_B \sum_n \frac{\partial p_n(B,T)}{\partial B} \left( \ln p_n(B,T) + 1 \right).
\end{aligned}
\end{equation}

Now, using \(\ln p_n(B,T) = -\beta E_n(B) - \ln Z\), we substitute to find,
\begin{equation}
\begin{aligned}
    \frac{\partial S(B,T)}{\partial B}
    &= -k_B \sum_n \frac{\partial p_n(B,T)}{\partial B} 
    \left( -\beta E_n(B) - \ln Z + 1 \right) \\
    &= \frac{1}{T} \sum_n \frac{\partial p_n(B,T)}{\partial B} E_n(B)
    + k_B (\ln Z - 1) \frac{\partial}{\partial B} \sum_n p_n(B,T).
\end{aligned}
\end{equation}

The second term vanishes since \(\sum_n p_n = 1\). Integrating both sides with respect to \(B\) from \(B_{\alpha}\) to \(B_{\beta}\), we get,
\begin{equation}
    \int_{B_{\alpha}}^{B_{\beta}} \frac{\partial S(B,T)}{\partial B} \, dB
    = \frac{1}{T} \sum_n \int_{B_{\alpha}}^{B_{\beta}} \frac{\partial p_n(B,T)}{\partial B} E_n(B) \, dB.
\end{equation}

Recognizing the right-hand side as Eq.~\eqref{eq:appendix D Q}, we arrive at:
\begin{equation}
    S(B_{\beta}, T) - S(B_{\alpha}, T) = \frac{Q}{T}
    \quad \Rightarrow \quad
    Q = T \big( S(B_{\beta}, T) - S(B_{\alpha}, T) \big).
    \label{eq:appendix d q2}
\end{equation}

Hence, Eq.~\eqref{eq: Heat_work} reduces to Eq.~\eqref{eq:appendix d q2} for an isothermal process.

\end{document}